\documentclass[amsmath, amssymb, superscriptaddress, reprint, colorlinks=true, citecolor=blue, linkcolor=blue, urlcolor=blue]{revtex4-1}

\usepackage[utf8]{inputenc}
\usepackage[T1]{fontenc}
\usepackage{newtxtext}
\usepackage{graphicx}
\usepackage{grffile}
\usepackage{longtable}
\usepackage{wrapfig}
\usepackage{rotating}
\usepackage[normalem]{ulem}
\usepackage{amsmath}
\usepackage{textcomp}
\usepackage{amssymb}
\usepackage{capt-of}
\usepackage{hyperref}
\usepackage{float}
\usepackage{bm}
\usepackage{enumitem}
\begin{document}

\title{
A kinetic--moment framework for electron energy dynamics in capacitively coupled plasmas: absorption, conversion, transport, and dissipation
}

\author{Jianxiong Yao}
\affiliation{School of Physics, Beijing Institute of Technology, Beijing 100081, People's Republic of China}

\author{Zeduan Zhang}
\affiliation{School of Physics, Beijing Institute of Technology, Beijing 100081, People's Republic of China}

\author{Feng He}
\affiliation{School of Physics, Beijing Institute of Technology, Beijing 100081, People's Republic of China}

\author{Jinsong Miao}
\affiliation{School of Physics, Beijing Institute of Technology, Beijing 100081, People's Republic of China}

\author{Jiting Ouyang}
\affiliation{School of Physics, Beijing Institute of Technology, Beijing 100081, People's Republic of China}

\author{Bocong Zheng}
\thanks{Author to whom correspondence should be addressed.}
\affiliation{School of Physics, Beijing Institute of Technology, Beijing 100081, People's Republic of China}

\date{\today; Email: zheng@bit.edu.cn [B. Zheng]}

\begin{abstract}
Understanding electron energy dynamics in low-temperature plasmas such as capacitively coupled plasmas (CCPs), including energy absorption, conversion, transport, and dissipation, is essential for interpreting discharge physics and process applications.
We propose a kinetic--moment framework based on particle-in-cell/Monte Carlo collision (PIC/MCC) simulations.
The framework reconstructs the first three velocity moments of the Boltzmann equation directly from PIC/MCC data and enables a quantitative, self-consistent description of electron energy dynamics in low-pressure CCPs.
To clarify energy conversion among electromagnetic energy, electron fluid kinetic (mechanical) energy, and electron thermal (internal) energy, we further separate the total energy transport equation into kinetic- and thermal-energy equations.
We find that, at low pressure, electrons gain directed kinetic energy in the sheath and convert it locally into thermal energy through pressure--strain interaction and collisions.
Thermal energy is then transported into the bulk and is dissipated mainly by inelastic electron--neutral collisions.
We further decompose pressure--strain interaction into reversible pressure dilatation and irreversible viscous-like dissipation, which correspond to conversion driven by volumetric compression or expansion and by shear deformation, respectively.
This decomposition reveals a significant thermalization channel beyond collisions.
More broadly, the results show coexistence of localized kinetic-to-thermal conversion near the sheath and nonlocal energy transport from the sheath to the bulk dominated by microscopic heat flux.
The heat flux deviates strongly from Fourier's law based on local temperature gradients.
This framework provides a clear fluid description with kinetic fidelity and offers a practical tool for analyzing energy evolution in nonequilibrium plasmas.
\end{abstract}

\maketitle

\section{Introduction}
\label{sec:introduction}

Low-temperature plasmas are nonlinear open systems far from thermodynamic equilibrium.
They are sustained by continuous external power input, which is coupled into the discharge through electromagnetic fields.
Energy absorbed by charged particles is redistributed by conversion and transport processes and is ultimately lost through boundary fluxes or through collisional dissipation within the discharge volume.
A quantitative description of electron energy dynamics, including absorption, conversion, transport, and dissipation, is essential for interpreting macroscopic plasma behavior and improving process performance~\cite{lieberman2005principles,li2018energy}.

Capacitively coupled plasmas (CCPs) driven by radio-frequency (RF) power are a typical low-temperature plasma source.
They are widely used in microelectronics manufacturing and biomedical applications, including semiconductor etching, thin-film deposition, and surface microstructuring or functionalization~\cite{lieberman2005principles,makabe2014plasma,pascal11_physic,chen2012introduction,bienholz2013multiple}.
In CCPs, electrons, ions, and neutrals have distinct energy scales and energy distributions.
Because of their small mass, electrons respond most strongly to the RF field and typically gain energy from the electric field more efficiently than ions.
Through collisions, electrons transfer energy to neutrals and sustain ionization and excitation processes.
In this context, the terms electron heating and electron power absorption are often used interchangeably, but they describe different aspects.
Electron power absorption refers to energy transfer from the electric field to electrons through directional acceleration, which drives an anisotropic velocity distribution.
Electron heating refers to the subsequent relaxation toward isotropy in velocity space, which manifests as an increase in electron temperature.
For clarity, we use electron power absorption to describe energy transfer from the electric field to electrons~\cite{schulze2018spatio,vass2020electron,wilczek2020electron}.

Ohmic power absorption due to electron--neutral collisions was first identified and dominates at relatively high pressures~\cite{lieberman2005principles,pascal11_physic}.
At low pressure, long mean free paths reduce the effectiveness of collisional absorption.
An additional collisionless, often termed stochastic, power absorption mechanism is therefore required to sustain the discharge~\cite{popov1985power,godyak1990abnormally}.
In the widely used hard-wall model, collisionless absorption is attributed to momentum exchange between electrons and oscillating sheaths~\cite{turner2009collisionless,lafleur2014equivalence}.
Total power absorption is therefore often decomposed into collisional (Ohmic) and collisionless (stochastic) contributions.
However, because these pictures rely on different assumptions, they do not form a unified and self-consistent framework, even though they can reproduce experimental trends in certain regimes~\cite{popov1985power,godyak1990abnormally}.

Surendra and Dalvie~\cite{surendra1993moment} were among the first to analyze electron power absorption in CCPs in a self-consistent manner by combining the Boltzmann equation with particle simulations.
Using PIC/MCC, they evaluated each term in the conservation equations directly, separated collisional and collisionless contributions, and tested common assumptions used in fluid models.
More than two decades later, Lafleur \emph{et al.}~\cite{lafleur2014electron} revisited electron power absorption using the electron momentum balance equation with more realistic cross sections.
They found that, in most cases, Ohmic heating and pressure-gradient-related pressure heating dominate the total absorption.
Building on this work, Schulze \emph{et al.}~\cite{schulze2018spatio} incorporated the continuity equation and decomposed electron power absorption into seven terms with clear physical interpretations.
This spatiotemporally resolved, self-consistent decomposition based on the momentum balance and the associated mechanical-energy equation is commonly referred to as Boltzmann term analysis~\cite{schulze2018spatio,vass2020electron}.
In recent years, this approach has been applied to a broad range of discharges, including low-pressure inert gases~\cite{wilczek2020electron,vass2020observation} and reactive gases~\cite{vass2020electron,proto2020electron,proto2021electron,derzsi2022electron}.
It has also been extended to magnetized CCP discharges~\cite{zheng2019enhancement,wang2020electron,zheng2021electron}, atmospheric-pressure microplasma jets~\cite{vass2021electron}, and Hall thrusters~\cite{lafleur2017role,faraji2022enhancing,charoy2020comparison}.

Most existing studies emphasize how electromagnetic fields transfer energy to charged particles to ignite and sustain the discharge.
Less attention has been paid to what happens afterward, namely how absorbed energy is converted, transported, and ultimately dissipated within the plasma~\cite{versolato2024plasma}.
These internal processes shape energy distributions and reaction rates and, therefore, strongly influence process performance.
In addition, Boltzmann term analysis is formulated mainly from the momentum balance and the associated mechanical-energy equation, whereas most electron energy in CCPs is thermal.
The field-work term \(\boldsymbol J \cdot \boldsymbol E\) quantifies how electromagnetic energy enters the charged-particle system, but it does not by itself show how that energy is redistributed in space and converted into thermal energy, that is, how electrons are thermalized.
Electron power absorption is therefore only the starting point of discharge energy dynamics.
Together with subsequent energy conversion, transport, and dissipation, it constitutes the full electron energy dynamics of the discharge.

At higher pressure, plasmas can often be treated as fluids with Maxwellian velocity distributions~\cite{lieberman2005principles,trelles2025fluid}.
In this regime, continuity, momentum, and energy equations can provide a useful macroscopic description.
Fluid models, however, face the moment-closure problem.
The number of unknowns exceeds the number of equations, so higher-order moments must be related to lower-order ones through closure assumptions.
This limitation becomes critical at low pressure, below a few tens of mTorr, which is relevant to many industrial applications.
Long mean free paths then prevent fast electrons from equilibrating rapidly.
This includes electrons accelerated by sheath expansion and secondary electrons that traverse the sheath and gain nearly the full sheath potential.
The discharge develops a substantial energetic population and the distribution function departs from a Maxwellian.
In this kinetic regime, conventional fluid models lose accuracy and do not capture energy conversion reliably.
Higher-moment fluid models can improve fidelity, but they still require self-consistent benchmarks, which are difficult to obtain within a purely fluid framework~\cite{miller2016multi,sahu2020full,kuldinow2024ten,park2024kinetic}.
Hybrid models that treat non-thermalized components with Monte Carlo methods can extend the range of fluid modeling to lower pressures~\cite{sommerer1992numerical,winske2003hybrid,kushner2009hybrid,wen2019secondary,kruger2021electric,zhang2023benchmarking}.
However, at sufficiently low pressure, fully kinetic approaches, such as PIC/MCC, are still required for self-consistent solutions and for benchmarking and validation~\cite{schulenberg2021multi,park2023experimental,simha2023kinetic}.

To address these issues, we propose a first-principles kinetic--moment framework.
Using PIC/MCC simulations, we access kinetic information directly and reconstruct the first three velocity moments of the Boltzmann equation without closure assumptions.
This approach combines the interpretability of moment-based transport analysis with fully kinetic fidelity.
Compared with Boltzmann term analysis, which focuses on the second velocity moment~\cite{surendra1993moment,lafleur2014electron,schulze2018spatio,zheng2019enhancement}, our framework extends moment-based diagnostics to the total energy transport equation involving the third velocity moment.
By separating kinetic and thermal energy dynamics, we establish a self-consistent description of electron energy dynamics that couples energy absorption, conversion, transport, and dissipation.

The remainder of this paper is organized as follows.
Section~\ref{sec:modeling} presents the theoretical framework and computational method, including the moment equations, the PIC/MCC model, and the simulation setup.
Section~\ref{sec:results} analyzes electron energy dynamics in low-pressure CCPs.
Subsection~\ref{sec:energyTransport} uses the total energy transport equation to characterize spatiotemporal transport and establish an overall view of the discharge.
Subsection~\ref{sec:kineticTransport} uses the decoupled kinetic-energy equation to study energy absorption, conversion, transport, and dissipation.
Subsection~\ref{subsec:BTAcomparison} compares kinetic-energy transport analysis with Boltzmann term analysis, and Subsection~\ref{subsec:pressureStrain} presents the pressure--strain decomposition and its physical interpretation.
Subsection~\ref{sec:thermalTransport} uses the decoupled thermal-energy equation to analyze thermal energy transport.
Section~\ref{sec:conclusions} summarizes the main findings and implications.

\section{THEORETICAL BACKGROUND AND COMPUTATIONAL METHOD}
\label{sec:modeling}

\subsection{Kinetic Moments of the Distribution Function and Transport Equation}
\label{sec:transportEquation}

Macroscopic transport in plasmas arises from microscopic particle motion that carries conserved quantities.
Because the number of particles is enormous and interactions include short-range collisions and long-range Coulomb forces, tracking individual trajectories is impractical.
In fluid models, a plasma is treated as an ensemble of charged fluids.
Its state is described by density, flow velocity, temperature, and related variables.
In the more fundamental kinetic description, the state of species \(s\) is specified by its phase-space distribution function \(f_s(\boldsymbol r,\boldsymbol v,t)\), and macroscopic quantities are obtained by velocity-space averaging.
The distribution function \(f_s(\boldsymbol r,\boldsymbol v,t)\) is the particle number density per unit phase-space volume at \((\boldsymbol r,\boldsymbol v)\) and time \(t\), so the number of particles in a phase-space element is
\begin{equation}
\mathrm{d}N_s = f_s(\boldsymbol r,\boldsymbol v,t)\,\mathrm d^3r\,\mathrm d^3v\,.
\end{equation}
For a microscopic quantity \(\chi(\boldsymbol r,\boldsymbol v,t)\), its macroscopic average is defined as
\begin{equation}
\langle\chi\rangle_s(\boldsymbol r,t)\equiv
\frac{\int \chi(\boldsymbol r,\boldsymbol v,t)\,f_s(\boldsymbol r,\boldsymbol v,t)\,\mathrm d^3v}{\int f_s(\boldsymbol r,\boldsymbol v,t)\,\mathrm d^3v}\,,
\end{equation}
where \(\langle\cdot\rangle\) denotes the weighted average over velocity space.
Specifically, macroscopic fluid variables correspond to velocity moments of \(f_s\).
The \(k\)th moment can be written as
\begin{equation}
\mathcal{M}_k(\boldsymbol r,t)=\int \underbrace{\boldsymbol v\boldsymbol v\cdots\boldsymbol v}_{k}\,f_s(\boldsymbol r,\boldsymbol v,t)\,\mathrm d^3v\,,
\end{equation}
which is a rank-\(k\) tensor.
The set \(\{\mathcal{M}_k\mid k=0,1,2,\dots\}\) provides an equivalent representation of the distribution function when sufficient smoothness is assumed~\cite{grad1949kinetic,braginskii1965transport,pitaevskii2012physical}.

Low-order moments have clear physical meaning.
The zeroth moment yields the number density,
\begin{equation}
n_s(\boldsymbol r,t)=\int f_s(\boldsymbol r,\boldsymbol v,t)\,\mathrm d^3v\,.
\end{equation}
The first moment yields the particle flux density,
\begin{equation}
\boldsymbol\Gamma_{\mathrm n,s}=n_s\boldsymbol u_s(\boldsymbol r,t)=\int \boldsymbol v\,f_s(\boldsymbol r,\boldsymbol v,t)\,\mathrm d^3v\,,
\end{equation}
which defines the flow velocity \(\boldsymbol u_s\).
Charge density \(\rho_{\mathrm c}=\sum_s q_s n_s\) and current density \(\boldsymbol J=\sum_s q_s n_s\boldsymbol u_s\) follow directly from the first two moments.
The second moment gives the momentum flux in the laboratory frame, also termed the stress tensor (momentum-flux dyad),
\begin{equation}
\boldsymbol{\mathcal T}_{\mathrm p,s}(\boldsymbol r,t)=m_s\int \boldsymbol v\boldsymbol v\,f_s(\boldsymbol r,\boldsymbol v,t)\,\mathrm d^3v\,.
\end{equation}
The third moment characterizes the energy flux density,
\begin{equation}
\boldsymbol\Gamma_{\varepsilon,s}(\boldsymbol r,t)=\frac12 m_s\int v^2\boldsymbol v\,f_s(\boldsymbol r,\boldsymbol v,t)\,\mathrm d^3v\,.
\end{equation}

The distribution function \(f_s\) evolves according to the Boltzmann equation,
\begin{equation}\label{eq:BoltzmannEq}
\frac{\partial f}{\partial t}+\boldsymbol v\cdot\nabla f+\boldsymbol a\cdot\nabla_v f=\left(\frac{\delta f}{\delta t}\right)_{\mathrm{coll}}\,,
\end{equation}
where \(\boldsymbol a\) denotes the acceleration due to external forces and the right-hand side accounts for collisions.
For brevity, we drop the species index \(s\) and consider only electromagnetic acceleration.
Multiplying Eq.~\eqref{eq:BoltzmannEq} by an arbitrary \(\chi(\boldsymbol r,\boldsymbol v,t)\) and integrating over velocity yields the general transport equation~\cite{bittencourt2013fundamentals},
\begin{equation}\label{eq:GeneralTransportEq}
\frac{\partial}{\partial t}\!\big(n\langle\chi\rangle\big)
+\nabla\!\cdot\!\big(n\langle\chi\,\boldsymbol v\rangle\big)
-n\big\langle\boldsymbol a\!\cdot\!\nabla_v\chi\big\rangle
=\left[\frac{\delta}{\delta t}\big(n\langle\chi\rangle\big)\right]_{\mathrm{coll}}\,.
\end{equation}
Equation~\eqref{eq:GeneralTransportEq} describes the spatiotemporal evolution of \(n\langle\chi\rangle\).
It contains four contributions, listed below.
\begin{enumerate}[leftmargin=*, label=(\roman*)]
\item \(\partial (n\langle\chi\rangle)/\partial t\) is the local time derivative at a fixed spatial position.
      In a periodic steady state, its average over one RF period is zero.
\item \(\nabla\!\cdot\!\big(n\langle\chi\,\boldsymbol v\rangle\big)\) is the divergence of the transport flux.
      It redistributes the conserved quantity in space.
      A positive divergence indicates net outflow and a negative divergence indicates net inflow.
      When integrated over the full volume, this term reduces to the net boundary flux by Gauss's theorem.
\item \(-n\langle\boldsymbol a\cdot\nabla_v\chi\rangle\) represents the effect of external forces.
      For \(\chi=m\boldsymbol v\) it becomes the Lorentz force \(\rho_{\mathrm c}\boldsymbol E+\boldsymbol J\times\boldsymbol B\). Assuming purely electrostatic acceleration (\(\boldsymbol B=0\)), this simplifies to \(\rho_{\mathrm c}\boldsymbol E\).
      For \(\chi=mv^2/2\) it becomes the field-work term \(\boldsymbol J\cdot\boldsymbol E\).
\item \(\left[\delta (n\langle\chi\rangle)/\delta t\right]_{\mathrm{coll}}\) is the collisional source or sink.
\end{enumerate}

Choosing \(\chi=m\), \(m\boldsymbol v\), and \(mv^2/2\) yields the mass, momentum, and energy transport equations, that is, the first three moment equations of the Boltzmann equation:
\begin{equation}\label{eq:TransportEq}
\begin{split}
&\frac{\partial\rho_{\mathrm m}}{\partial t}+\nabla\!\cdot\!\boldsymbol\Gamma_{\mathrm m}
=\left(\frac{\delta\rho_{\mathrm m}}{\delta t}\right)_{\mathrm{coll}}\,,\\
&\frac{\partial\boldsymbol\rho_{\mathrm p}}{\partial t}+\nabla\!\cdot\!\boldsymbol{\mathcal T}_{\mathrm p}-\rho_{\mathrm c}\boldsymbol E
=\left(\frac{\delta\boldsymbol\rho_{\mathrm p}}{\delta t}\right)_{\mathrm{coll}}\,,\\
&\frac{\partial\rho_\varepsilon}{\partial t}+\nabla\!\cdot\!\boldsymbol\Gamma_\varepsilon-\boldsymbol J\cdot\boldsymbol E
=\left(\frac{\delta\rho_\varepsilon}{\delta t}\right)_{\mathrm{coll}}\,.
\end{split}
\end{equation}
Here, \(\rho_{\mathrm m}=mn\), \(\boldsymbol\rho_{\mathrm p}=\rho_{\mathrm m}\boldsymbol u\), and \(\rho_\varepsilon=\frac12 m\int v^2 f(\boldsymbol x,\boldsymbol v,t)\,\mathrm d^3v\) are the mass, momentum, and total energy densities.
The corresponding collision terms represent their rates of change due to collisions.
In the divergence terms, \(\boldsymbol\Gamma\) or \(\boldsymbol{\mathcal T}\) denotes the flux of the corresponding conserved quantity per unit area and unit time.
The mass flux \(\boldsymbol\Gamma_{\mathrm m}=m\boldsymbol\Gamma_{\mathrm n}=\rho_{\mathrm m}\boldsymbol u\) is equivalent to the momentum density and is a vector.
The momentum flux \(\boldsymbol{\mathcal T}_{\mathrm p}\) describes transport of momentum components across different directions and the associated shear stresses, therefore it is a second-order tensor (a dyad).
The energy flux \(\boldsymbol\Gamma_\varepsilon\) is a vector and will be discussed below.
Dimensional analysis suggests that a flux is often written as a product of a density and a flow velocity.
In kinetic theory, however, flux moments mix bulk drift and thermal motion.
It is therefore useful to separate convective transport due to drift from non-convective transport due to random motion, which appears as stress and heat conduction.

The physical meaning of these moments is clearest in the local rest frame, where \(\boldsymbol u=0\) and the peculiar velocity is \(\boldsymbol w=\boldsymbol v-\boldsymbol u\).
In this frame, the momentum flux reduces to the pressure tensor and the energy flux reduces to the heat flux:
\begin{equation}
\begin{alignedat}{2}
\boldsymbol{\mathcal P}
&= m\int \boldsymbol w\boldsymbol w\,f(\boldsymbol r,\boldsymbol v,t)\,\mathrm d^3v
&={}&\boldsymbol{\mathcal T}_{\mathrm p}-\rho_{\mathrm m}\boldsymbol u\boldsymbol u\,,\\
\boldsymbol q
&=\tfrac12 m\int w^2\boldsymbol w\,f(\boldsymbol r,\boldsymbol v,t)\,\mathrm d^3v
&={}&\boldsymbol\Gamma_\varepsilon-\boldsymbol{\mathcal P}\!\cdot\!\boldsymbol u-\rho_\varepsilon\boldsymbol u\,.
\end{alignedat}
\end{equation}
The (thermal) pressure tensor \(\boldsymbol{\mathcal P}\) characterizes momentum transport due to random thermal motion, while the heat flux \(\boldsymbol q\) represents energy transport carried by microscopic thermal motion.
The moments measured in the laboratory frame and in the rest frame are related by Galilean transformation.
Substitution gives~\cite{bittencourt2013fundamentals}
\begin{equation}
\begin{alignedat}{3}
&\boldsymbol\Gamma_{\mathrm m} &&=\rho_{\mathrm m}\langle\boldsymbol v\rangle &&=\rho_{\mathrm m}\boldsymbol u\,,\\
&\boldsymbol{\mathcal T}_{\mathrm p} &&=\rho_{\mathrm m}\langle\boldsymbol v\boldsymbol v\rangle &&=\boldsymbol\rho_{\mathrm p}\boldsymbol u+\boldsymbol{\mathcal P}\,,\\
&\boldsymbol\Gamma_\varepsilon &&=\tfrac12\rho_{\mathrm m}\langle v^2\boldsymbol v\rangle &&=\rho_\varepsilon\boldsymbol u+\boldsymbol{\mathcal P}\cdot\boldsymbol u+\boldsymbol q\,.
\end{alignedat}
\end{equation}

The total fluid energy density can be decomposed into directed kinetic energy (also called drift kinetic energy) associated with bulk drift and thermal energy associated with random motion.
The kinetic-energy density associated with bulk drift is \(\rho_{\varepsilon,\mathrm k}=\rho_{\mathrm m}u^2/2\), which is also called the mechanical-energy density and equals the dynamic pressure.
The thermal-energy density is \(\rho_{\varepsilon,\mathrm{th}}=\frac12 m\int \boldsymbol w^2 f(\boldsymbol r,\boldsymbol v,t)\,\mathrm d^3v=\frac32 p\), which corresponds to the internal energy density associated with random motion, and equals the static pressure.
Here \(p=\frac13\mathrm{Tr}(\boldsymbol{\mathcal P})\) is the scalar isotropic pressure, defined as one third of the trace of the pressure tensor.
Starting from the total energy transport equation and using the energy-flux decomposition, one can derive separate transport equations for kinetic and thermal energy.
They read
\begin{widetext}
\begin{equation}
\begin{alignedat}{9}
& \frac{ \partial \rho_{\varepsilon,\mathrm k} }{ \partial t }
& + \nabla\!\cdot\!( \rho_{\varepsilon,\mathrm k} \boldsymbol u )
& + \nabla\!\cdot\!( \boldsymbol{\mathcal P} \cdot \boldsymbol u )
& - ( \boldsymbol{\mathcal P} \cdot \nabla ) \cdot \boldsymbol u
& - \boldsymbol J \cdot \boldsymbol E
& =
&
& \boldsymbol u \cdot \left( \frac{ \delta \boldsymbol \rho_\mathrm p }{ \delta t } \right)_\mathrm{coll}
& - \frac{1}{2} \left( \frac{ \delta \rho_\mathrm m }{ \delta t } \right)_\mathrm{coll} u^2 \,, \\
& \frac{ \partial \rho_{\varepsilon,\mathrm{th}} }{ \partial t }
& + \nabla\!\cdot\!( \rho_{\varepsilon,\mathrm{th}} \boldsymbol u )
&
& + ( \boldsymbol{\mathcal P} \cdot \nabla ) \cdot \boldsymbol u
& + \nabla\!\cdot\!\boldsymbol q
& = & \left( \frac{ \delta \rho_\varepsilon }{ \delta t } \right)_\mathrm{coll} -
& \boldsymbol u \cdot \left( \frac{ \delta \boldsymbol \rho_\mathrm p }{ \delta t } \right)_\mathrm{coll}
& + \frac{1}{2} \left( \frac{ \delta \rho_\mathrm m }{ \delta t } \right)_\mathrm{coll} u^2 \,.
\end{alignedat}
\end{equation}
\end{widetext}

\subsection{Model Description and Computational Implementation}
\label{sec:dischargeSetup}

A kinetic description of transport in CCPs requires access to the distribution function \(f(\boldsymbol r,\boldsymbol v,t)\).
Fluid models usually solve only a truncated set of moment equations of the Boltzmann or Vlasov equation and evolve macroscopic variables such as density, mean velocity, and energy.
In principle, retaining the full hierarchy of moments would recover the distribution function.
In practice, the moment-closure problem forces truncation at finite order and introduces constitutive closures such as Newtonian viscosity and Fourier's law.
Such closures typically assume near-local equilibrium.
This assumption breaks down at low pressure, where strong nonequilibrium features, such as energetic tails and anisotropy, are present.

PIC/MCC simulations provide a self-consistent kinetic description and are well suited for studying transport of charged particles in CCPs.
Rather than solving the distribution function or its moments directly, PIC/MCC represents \(f\) statistically by a large number of superparticles and follows their phase-space evolution~\cite{donko2011particle,donko2021edupic}.
Particles evolve under the self-consistent electric field, while collisions with the background gas are treated by Monte Carlo methods~\cite{vahedi1995monte}.
By collecting local velocity statistics, one can reconstruct the distribution function.
Moments of any order can then be obtained.
In practice, field quantities such as electric field, charge density, and particle density are defined on grid points.
Reconstructing \(f\) everywhere and recomputing moments in post-processing is redundant and inefficient.
Instead, the required moment quantities can be diagnosed directly by sampling particle data on the same spatial and temporal grids.
Collisional contributions can be obtained from particle property changes across each MCC step.
We do not repeat implementation details here.
Similar approaches are described in Refs.~\cite{lafleur2014electron,wu2022note}.
In summary, PIC/MCC provides macroscopic quantities with kinetic fidelity.
These quantities can be inserted directly into the velocity-moment equations, which enables reconstruction of transport equations without closure assumptions.

We use our in-house ASTRA code to perform 1d3v (one-dimensional in configuration space and three-dimensional in velocity space) PIC/MCC simulations of a geometrically symmetric argon CCP~\cite{zheng2019enhancement,zheng2021electron,yao2025unified}.
The discharge is formed between two parallel-plate electrodes separated by \(L_{\mathrm{gap}}=25~\mathrm{mm}\).
One electrode is driven by \(\phi=\phi_0\cos(2\pi f t)\) with \(f=27.12~\mathrm{MHz}\) and \(\phi_0=500~\mathrm{V}\), and the other electrode is grounded.
The neutral argon background is uniform with pressure \(p_{\mathrm g}=5~\mathrm{Pa}\) and temperature \(T_{\mathrm g}=300~\mathrm{K}\).
The simulation includes electrons and \(\mathrm{Ar}^+\) ions.
Electron--neutral collisions include elastic scattering, excitation, and ionization.
Ion--neutral collisions include isotropic scattering and backscattering.
Cross sections are taken from Ref.~\cite{surendra1990self}.
To focus on basic transport physics, particle reflection and secondary electron emission at the electrodes are neglected.

We choose this low-pressure case because the electron mean free path exceeds the characteristic system scale.
Transport then becomes strongly nonlocal.
It is not controlled solely by local electric fields or local gradients of density and temperature.
Instead, it depends on conditions over a broader spatial region.
This nonlocality is a key reason why local-field approximations in fluid models fail.
The present case therefore serves as a useful benchmark for kinetic studies of nonlocal transport.
All results are obtained using an implicit algorithm with an energy-conserving scheme.
We use 512 spatial cells and resolve each RF period with 1600 time steps.
After reaching periodic steady state, we track about \(4\times10^5\) superparticles and average over about 2000 RF periods to reduce statistical noise.
The simulation uses the electrostatic approximation.
Each transport term is diagnosed independently.
Consistency and completeness can be confirmed by comparing the sum of terms on both sides of each transport equation.

\section{RESULTS AND DISCUSSION}
\label{sec:results}

\subsection{Total Energy Transport Analysis}
\label{sec:energyTransport}

In a steady-state CCP, conserved quantities reach a periodic balance and repeat from one RF period to the next.
To identify the mechanisms responsible for temporal modulation and spatial redistribution, we analyze each term in the total energy transport equation and in the decoupled kinetic- and thermal-energy equations.
We also present time-averaged spatial profiles and space-averaged temporal evolution over one RF period.
Together, these diagnostics provide a self-consistent picture of electron energy balance and transport.

Wu \emph{et al.}~\cite{wu2022note} used PIC/MCC to compute terms in the energy balance of electromagnetic fields and charged particles in a CCP, reporting energy densities, fluxes, absorption, and dissipation.
However, they did not present the standard moment-equation form of the particle energy transport equation or discuss the divergence of particle energy flux, both of which are central for interpreting spatial redistribution.
Here, as described in Subsection~\ref{sec:transportEquation}, we derive the standard energy transport equation from velocity moments and further decouple it into kinetic- and thermal-energy transport equations.

For clarity, we rewrite the total energy transport equation by placing the source term on the left-hand side~\cite{yao2025similarity}.
\begin{equation}\label{eq:EnergyTransportEq}
\underbrace{\boldsymbol J\cdot\boldsymbol E}_{1}
=
\underbrace{\frac{\partial \rho_\varepsilon}{\partial t}}_{2}
\;+\;
\underbrace{\nabla\!\cdot\!\boldsymbol{\Gamma}_{\varepsilon}}_{3}
\;\underbrace{\mathrel{-}\,\left(\frac{\delta \rho_\varepsilon}{\delta t}\right)_{\mathrm{coll}}}_{4}\,.
\end{equation}
Here, \(\rho_\varepsilon\) is the total energy density, \(\boldsymbol\Gamma_\varepsilon\) is the energy flux, and \(\boldsymbol J\) and \(\boldsymbol E\) are the current density and electric field.
The terms in Eq.~\eqref{eq:EnergyTransportEq} have the following meanings.
\begin{enumerate}[leftmargin=*]
\item \(\boldsymbol J\cdot\boldsymbol E\) is the power density absorbed from the electromagnetic field.
      It is the source term.
\item \(\partial\rho_\varepsilon/\partial t\) is the local time variation of energy density.
      Its average over one RF period is zero in a periodic steady state.
\item \(\nabla\!\cdot\!\boldsymbol\Gamma_\varepsilon\) is the divergence of energy flux, which represents spatial redistribution.
      A positive value indicates net outflow and a negative value indicates net inflow.
      By Gauss's theorem, boundary losses can be evaluated by integrating the energy flux over electrode surfaces, \(L_{\varepsilon,\mathrm{loss}}^{\mathrm{boundary}}=\oint_{A_{\mathrm{electrode}}}\boldsymbol\Gamma_\varepsilon\cdot\mathrm d\boldsymbol A=\iiint_V(\nabla\cdot\boldsymbol\Gamma_\varepsilon)\,\mathrm dV\).
      The same idea applies to other divergence terms.
\item \(-(\delta\rho_\varepsilon/\delta t)_{\mathrm{coll}}\) is the collisional rate of change of energy density.
      In simple discharges where super-elastic processes are negligible, it is positive and represents net energy transfer from charged particles to the neutral gas through heating, excitation, and ionization.
\end{enumerate}

\begin{figure*}[htbp]
\centering
\includegraphics[width=0.8\textwidth]{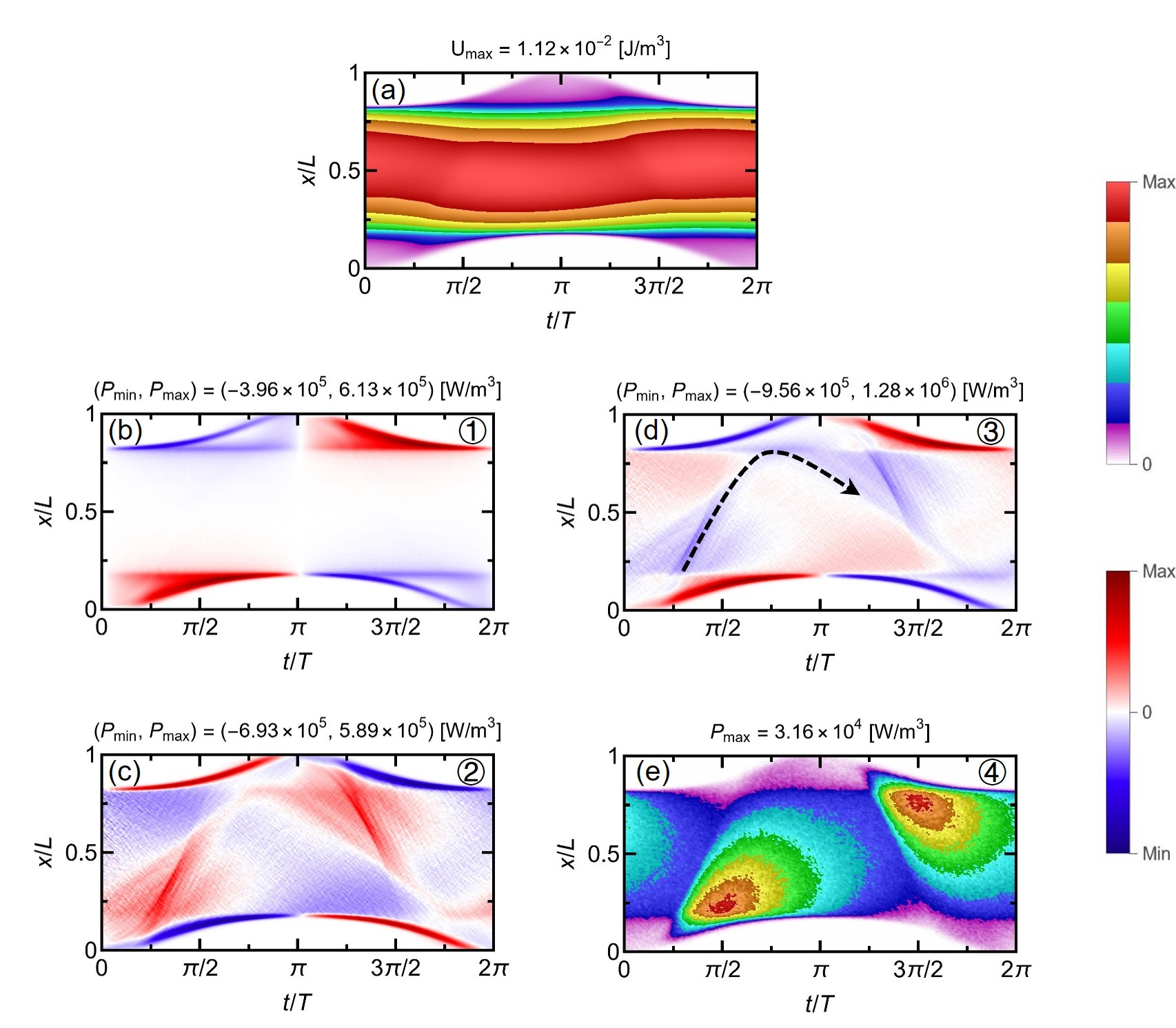}
\caption{\label{fig:energy}Spatiotemporal distributions of energy density \(\rho_\varepsilon\) (a) and each transport term (b-d) corresponding to energy transport equation~\eqref{eq:EnergyTransportEq} for electrons.}
\end{figure*}

\begin{figure*}[htbp]
\centering
\includegraphics[width=0.8\textwidth]{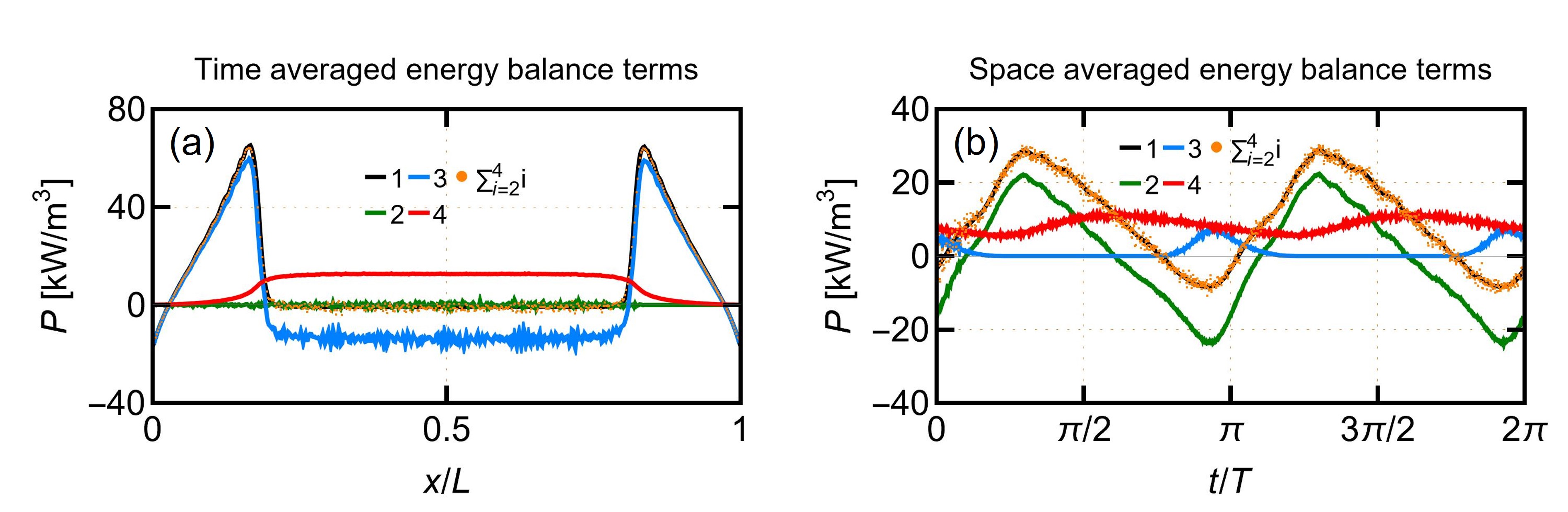}
\caption{\label{fig:energyAvg}Time-averaged spatial profile (a) and space-averaged temporal evolution (b) of energy transport terms for electrons.
         The numbered labels in the figure correspond to the terms in energy transport equation~\eqref{eq:EnergyTransportEq}.}
\end{figure*}

These terms quantify distinct physical processes, and their relative contributions provide insight into electron energy dynamics.
Figures~\ref{fig:energy} and~\ref{fig:energyAvg} show spatiotemporal distributions of electron energy density and the corresponding transport terms over one RF period, together with time-averaged spatial profiles and space-averaged temporal evolution.
In Fig.~\ref{fig:energyAvg}, the left-hand side of Eq.~\eqref{eq:EnergyTransportEq} agrees with the sum of the right-hand side terms for both the time-averaged and space-averaged diagnostics.
This agreement confirms the self-consistency and accuracy of the simulation and the moment-based diagnostics.

Energy transfer to charged particles originates from work done by electromagnetic fields on the electron and ion fluids.
Under RF driving, a CCP behaves as a nonlinear capacitor that stores and releases electromagnetic energy within each RF period.
A large fraction of the injected energy is stored in the fields and returned to the external circuit.
Energy exchange between fields and particles occurs through acceleration and deceleration in the electric field, and only a small fraction of the injected RF energy is converted into particle energy as net field work, \(\boldsymbol J\cdot\boldsymbol E\).
Here, we focus on how particle energy is transported after absorption.
For a discussion of electromagnetic energy transport, see Ref.~\cite{wu2022note}.

Figure~\ref{fig:energy}(a) shows the spatiotemporal distribution of total electron energy density.
Similar to the electron density profile, it is mainly concentrated in the bulk.
It is also modulated in time and space due to periodic RF power injection and sheath motion.
Because the applied field is screened over the sheath scale, the bulk electric field is weak.
Electrons therefore absorb energy mainly near the sheath region.
Time-averaged results in Fig.~\ref{fig:energyAvg}(a) show that the source term \(\boldsymbol J\cdot\boldsymbol E\) is localized in the sheath region, defined by the maximum sheath width, and is nearly zero in the bulk.
Space-averaged results in Fig.~\ref{fig:energyAvg}(b) show that \(\boldsymbol J\cdot\boldsymbol E\) is strongly modulated by the time-dependent sheath.
Electrons absorb energy mainly during sheath expansion and lose energy during sheath collapse.
The net energy gain over a full RF period is positive.
This confirms that energy exchange between the electromagnetic field and the electron fluid is bidirectional.
As expected, the period-averaged contribution of \(\partial\rho_\varepsilon/\partial t\) vanishes at every spatial position, because density and temperature do not change from one RF period to the next in steady state.
This means that energy gained from the field must be balanced by boundary losses or collisional dissipation.
These correspond to the third and fourth terms in Eq.~\eqref{eq:EnergyTransportEq}.
If the discharge has not reached steady state, the time-variation term does not average to zero.
It can therefore be used as a criterion for steady state~\cite{chen2023note}.
The space-averaged evolution in Fig.~\ref{fig:energyAvg}(b) shows that the instantaneous time-variation term nearly follows the source term, which indicates that local energy density changes are mainly driven by field work.

The divergence term \(\nabla\!\cdot\!\boldsymbol\Gamma_\varepsilon\) in Fig.~\ref{fig:energy}(d) is nearly the negative of the time-variation term in Fig.~\ref{fig:energy}(c).
Near the sheath, where energy absorption is strongest, the divergence term is positive.
This indicates rapid energy outflow from the region and a sharp local decrease in energy density.
Physically, this corresponds to energetic electron beams produced by sheath expansion.
These beams carry energy across the gap and generate a strong energy flux that transports energy from the sheath into the bulk.
In Fig.~\ref{fig:energy}(d), a clear energy-transport path crosses the bulk and is marked by a black dashed line.
It represents an energetic electron beam with a well-defined propagation direction.
The electrons originate from acceleration by the expanding sheath at one electrode.
They traverse the entire gap and interact with the opposite sheath during its collapse phase.
Part of the beam is reflected and returns to the bulk.
This behavior is common at low pressure when the mean free path exceeds the electrode gap.
Energetic electrons then propagate almost collisionlessly and interact with the opposite sheath.
Although they constitute only a small fraction of the population, about 1\%, they are responsible for ionization and play a major role in excitation.
Their kinetics and energy distribution depend strongly on sheath dynamics and mean free path effects, including sheath-edge expansion velocity and sheath electric fields.
Experiments and simulations show that basic discharge parameters such as gap size, driving frequency, and voltage waveform can control these energetic beams~\cite{yao2025unified,liu2011collisionless,wilczek2015effect}.
Recent studies~\cite{heil2008numerical,schulze2008stochastic,schulze2010phase,bruneau2015strong} link pronounced ionization or excitation peaks to energetic electron beams launched from expanding sheaths.
However, without a direct diagnostic, such interpretations can miss details or lead to ambiguity.
The divergence of the energy flux provides a direct measure that resolves energetic-electron trajectories and their interaction with sheaths.

Time-averaged results in Fig.~\ref{fig:energyAvg}(a) show this nonlocality more clearly.
The flux-divergence term overlaps the energy source term in the sheath and takes a nearly constant negative value in the bulk that opposes the collisional loss term.
This indicates that energy is absorbed in the sheath, redistributed by energy flux divergence, and then dissipated in the bulk by collisions.
It confirms that, in low-pressure CCPs, energy absorption in the sheath and dissipation in the bulk are separated in space.
Space-averaged results in Fig.~\ref{fig:energyAvg}(b) show that the divergence term becomes positive only during sheath collapse.
This indicates that electrons reach the electrodes and carry energy out of the system.
This boundary loss is small.
The divergence term mainly redistributes energy in space.
Energy is dissipated mainly through collisions throughout the RF period, and its phase lags the source term.
This phase lag reflects the time required for energy transport and deposition.
Electrons generally travel from the sheath to the bulk before most of their energy is dissipated through collisions with neutrals.
In low-pressure CCPs with strong secondary electron emission, the contribution of the flux-divergence term is expected to increase~\cite{lisovskiy2004alpha,wen2019secondary,wen2023field,noesges2023nonlocal}.

The total energy transport equation captures energy absorption, transport, and dissipation, but it does not distinguish among different energy forms.
As noted above, the total energy density contains the kinetic energy density \(\rho_{\varepsilon,\mathrm k}=\rho_{\mathrm m}u^2/2\) and the thermal energy density \(\rho_{\varepsilon,\mathrm{th}}=3p/2\).
We therefore decouple the total energy equation into kinetic and thermal energy transport equations.
This enables separate analysis of their transport behaviors and of conversion between them.

\subsection{Kinetic Energy Transport Analysis}
\label{sec:kineticTransport}

In low-pressure CCPs, electron fluid kinetic energy corresponds to directed mechanical motion. The generation and dissipation of this energy are pivotal for electron power absorption and thermalization.
In this subsection, we use the kinetic-energy transport equation to analyze conversion among electromagnetic energy, electron kinetic energy, and electron thermal energy, together with spatial redistribution of kinetic energy.

To highlight energy conversion, we group on the left-hand side the terms that change the energy form, including the source that converts electromagnetic energy into kinetic energy and the sinks that convert kinetic energy into thermal energy.
We keep the conservative time-derivative and divergence terms on the right-hand side.
The kinetic-energy transport equation can be written as
\begin{widetext}
\begin{equation}\label{eq:KineticTransportEq}
\underbrace{\boldsymbol J\cdot\boldsymbol E}_{1}
\;+\;
\underbrace{(\boldsymbol{\mathcal P}\!\cdot\!\nabla)\!\cdot\!\boldsymbol u}_{2}
\;\underbrace{\mathrel{-}\,\frac12\!\left(\frac{\delta \rho_{\mathrm m}}{\delta t}\right)_{\mathrm{coll}} u^{2}}_{3}
\;+\;
\underbrace{\boldsymbol u \cdot \left(\frac{\delta \boldsymbol{\rho}_{\mathrm p}}{\delta t}\right)_{\mathrm{coll}}}_{4}
=
\underbrace{\frac{\partial \rho_{\varepsilon,\mathrm k}}{\partial t}}_{5}
\;+\;
\underbrace{\nabla\!\cdot\!\big(\rho_{\varepsilon,\mathrm k}\,\boldsymbol u\big)}_{6}
\;+\;
\underbrace{\nabla\!\cdot\!\big(\boldsymbol{\mathcal P}\cdot \boldsymbol u\big)}_{7}\,.
\end{equation}
\end{widetext}
Here, \(\rho_{\varepsilon,\mathrm k}=\rho_{\mathrm m}u^2/2\) is the fluid kinetic energy density, \(\boldsymbol u\) is the macroscopic flow velocity, and \(\boldsymbol{\mathcal P}\) is the pressure tensor.
Terms 5--7 are conservative.
They redistribute the kinetic-energy density in space and time through transport.
As conservative terms, they do not change the energy form or the total kinetic energy in the domain, except for boundary fluxes that carry kinetic energy out of the system.
We next summarize the meaning of each term.
\begin{enumerate}[leftmargin=*]
\item \(\boldsymbol J\cdot\boldsymbol E\) is the work done by the electric field on the electron fluid.
      It converts electromagnetic energy into directed kinetic energy and is the kinetic-energy source.
\item \((\boldsymbol{\mathcal P}\!\cdot\!\nabla)\!\cdot\!\boldsymbol u\) represents the effect of the pressure tensor acting on velocity gradients.
      It describes bidirectional conversion between kinetic and thermal energy caused by volumetric compression or expansion and shear deformation of fluid elements.
      It is commonly called the pressure--strain interaction and provides a channel for kinetic-to-thermal conversion~\cite{cassak2022pressure,cassak2022pressureII,barbhuiya2022pressure}.
\item \(-\frac12(\delta\rho_{\mathrm m}/\delta t)_{\mathrm{coll}}u^2\) is the kinetic energy change associated with mass change due to collisions.
      When ionization produces charged particles, \((\delta\rho_{\mathrm m}/\delta t)_{\mathrm{coll}}>0\), energy is required to accelerate newly created particles to the local drift velocity.
      This makes the term a kinetic-energy sink.
      Loss processes such as recombination are absent in the present conditions.
\item \(\boldsymbol u\cdot(\delta\boldsymbol\rho_{\mathrm p}/\delta t)_{\mathrm{coll}}\) is the kinetic energy change due to collisional momentum transfer.
      It represents irreversible conversion of directed kinetic energy into random thermal energy through momentum scattering with neutrals.
      It is a major dissipation mechanism.
\item \(\partial\rho_{\varepsilon,\mathrm k}/\partial t\) is the local time variation of kinetic energy density and averages to zero in steady state.
\item \(\nabla\!\cdot(\rho_{\varepsilon,\mathrm k}\boldsymbol u)\) is convective transport of kinetic energy by bulk motion.
\item \(\nabla\!\cdot(\boldsymbol{\mathcal P}\cdot\boldsymbol u)\) is kinetic-energy transport associated with pressure work.
      It describes redistribution of kinetic energy due to spatially varying pressure.
\end{enumerate}

\begin{figure*}[htbp]
\centering
\includegraphics[width=0.8\textwidth]{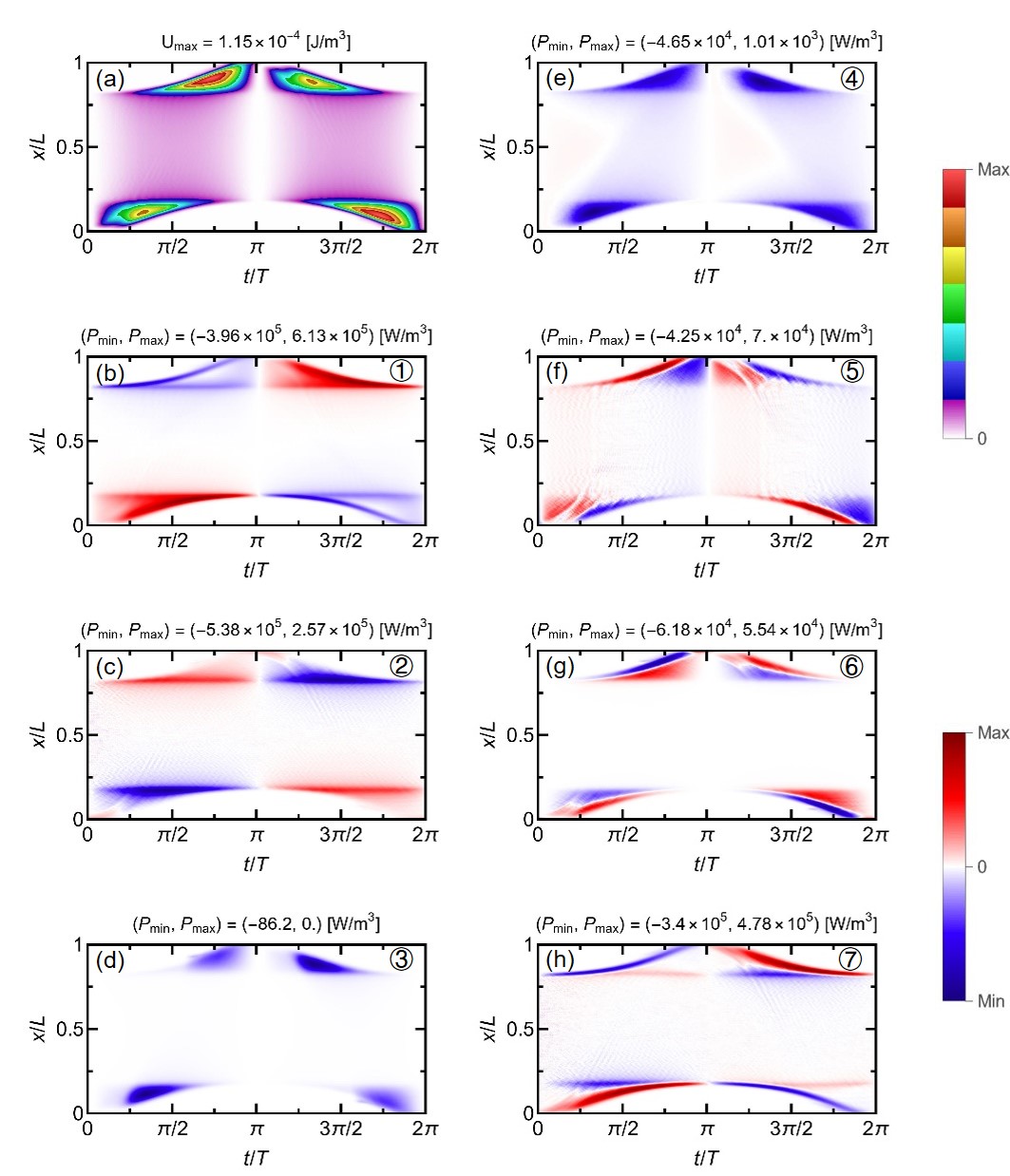}
\caption{\label{fig:kinetic}Spatiotemporal distributions of kinetic energy density \(\rho_{\varepsilon, \mathrm k}\) (a) and each transport term (b-h) corresponding to kinetic energy transport equation~\eqref{eq:KineticTransportEq} for electrons.}
\end{figure*}

\begin{figure*}[htbp]
\centering
\includegraphics[width=0.8\textwidth]{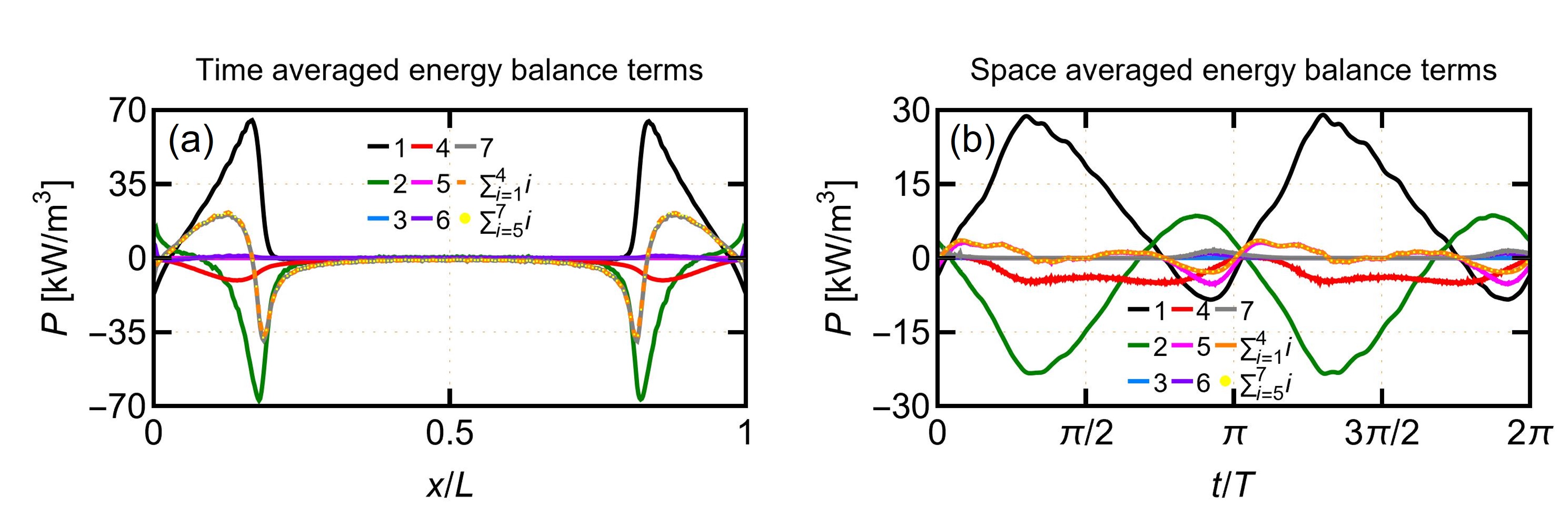}
\caption{\label{fig:kineticAvg}Time-averaged spatial profile (a) and space-averaged temporal evolution (b) of kinetic energy transport terms for electrons.
         The numbered labels in the figure correspond to the terms in kinetic energy transport equation~\eqref{eq:KineticTransportEq}.}
\end{figure*}

Equation~\eqref{eq:KineticTransportEq} provides a local balance for electron fluid kinetic energy and shows that conversion into thermal energy is the primary dissipation pathway.
Figure~\ref{fig:kinetic} shows spatiotemporal distributions of kinetic energy density and each term in Eq.~\eqref{eq:KineticTransportEq}.
Figure~\ref{fig:kineticAvg} shows time-averaged spatial profiles and space-averaged temporal evolution.

As noted above, \(\boldsymbol J\cdot\boldsymbol E\) is the only mechanism by which charged particles gain energy, representing acceleration by the electric field.
In this process, electromagnetic energy is converted into directed kinetic energy \(\rho_{\varepsilon,\mathrm k}\), which corresponds to an increase in drift speed.
This conversion occurs mainly in the sheath region, due to the strong sheath field and the ambipolar field near the maximum sheath width.
In contrast, the bulk field is weak at low pressure~\cite{schulze2011ionization,schulze2014effect}.
As discussed in Subsection~\ref{sec:energyTransport}, electron power absorption oscillates in time with sheath expansion and collapse.
When the sheath nearly collapses on one side, \(\boldsymbol J\cdot\boldsymbol E\) can become negative.
This indicates a return of energy from electrons to the field and reflects bidirectional coupling between field energy and particle kinetic energy.

Figure~\ref{fig:kineticAvg} shows that the pressure--strain term follows a trend opposite to \(\boldsymbol J\cdot\boldsymbol E\), both in the space-averaged time evolution and in the time-averaged spatial profile.
It is also localized in the sheath.
In the expansion phase, for example during the first quarter of an RF period, kinetic energy increases while the pressure--strain term becomes strongly negative in nearly the same region.
This indicates conversion from directed kinetic energy into thermal energy.
During sheath collapse, the pressure--strain term can reverse sign and convert thermal energy back to directed kinetic energy, which can then be returned to the field.
The pressure--strain interaction therefore provides a bidirectional channel between directed kinetic energy and thermal energy.
The underlying conversion mechanisms are discussed in Subsection~\ref{subsec:pressureStrain}.

The third and fourth terms on the left-hand side of Eq.~\eqref{eq:KineticTransportEq} are collisional terms.
Together they describe conversion of directed kinetic energy into thermal energy through collisional changes in momentum and mass.
The drag term \(\boldsymbol u\cdot(\delta\boldsymbol\rho_{\mathrm p}/\delta t)_{\mathrm{coll}}\) is the dominant collisional dissipation channel.
When electron speeds are much larger than neutral speeds and elastic collisions dominate, the collisional momentum loss rate along the field direction can be written as \(-(\delta\rho_{\mathrm p,ex}/\delta t)_{\mathrm{coll}}=m_{\mathrm e}n_{\mathrm g}\int\sigma_{\mathrm m}|\boldsymbol v_{\mathrm e}|v_{\mathrm e x}f_{\mathrm e}\,\mathrm d^3v\), where \(n_{\mathrm g}\) is the gas density and \(\sigma_{\mathrm m}\) is the momentum-transfer cross section~\cite{robson2006introductory}.
Macroscopically, this term acts as friction and opposes the drift.
In the absence of inelastic processes, elastic collisions remove little kinetic energy per collision, but they continually randomize the velocity direction, that is, isotropize the distribution.
This ongoing randomization transfers the directed kinetic energy associated with the drift velocity \(\boldsymbol u\) into random thermal energy.
Time-averaged results show that collisional kinetic-energy loss is concentrated near the sheath, coinciding with the field-to-kinetic conversion region, and is negligible in the bulk.
As shown in Fig.~\ref{fig:kinetic}(a), high drift energy near the sheath during both expansion and collapse makes each collision more effective at scattering momentum, so dissipation peaks there.
As indicated in Fig.~\ref{fig:energy}(d) in Subsection~\ref{sec:energyTransport}, energetic electrons originating from the opposite expanding sheath traverse the bulk to interact with the local sheath during its collapse.
Consequently, both phases have similarly high drift energy near the sheath, rendering collisional thermalization approximately symmetric between expansion and collapse.
Under the present low-pressure conditions, infrequent collisions make pressure--strain interaction the dominant pathway for converting kinetic energy into thermal energy; at higher pressure, the drag term is expected to dominate~\cite{lafleur2014electron,vass2022evolution}.

Figures~\ref{fig:kinetic} also show that the conservative terms on the right-hand side of Eq.~\eqref{eq:KineticTransportEq} are localized in the sheath.
Most kinetic energy gained in the sheath is converted locally into thermal energy through pressure--strain interaction and collisional friction.
The remaining kinetic energy is only a tiny fraction of the total and is lost as kinetic-energy flux during sheath collapse.
Because kinetic-energy transport is confined to the sheath, its contribution to global energy redistribution is negligible.
The convective transport term in Fig.~\ref{fig:kinetic}(g) averages to nearly zero at all positions, while the pressure-driven kinetic-energy transport in Fig.~\ref{fig:kinetic}(h) dominates redistribution of the small amount of kinetic energy that remains.
Global energy redistribution in the discharge is therefore governed mainly by thermal energy transport.
Figure~\ref{fig:kineticAvg} shows good agreement between the summed left- and right-hand sides of Eq.~\eqref{eq:KineticTransportEq}, confirming kinetic-energy conservation and the numerical accuracy of the diagnostics.

\subsection{Comparison with Boltzmann Term Analysis}
\label{subsec:BTAcomparison}

Boltzmann term analysis (BTA) is a standard approach for diagnosing electron power absorption in CCPs.
It is rooted in the momentum balance and the associated mechanical-energy equation.
Here we relate BTA to the kinetic-energy transport analysis by deriving the BTA form and mapping its terms onto Eq.~\eqref{eq:KineticTransportEq}.

In BTA, the total electric field \(\boldsymbol E\) can be viewed as a superposition of ``effective fields'' that balance other momentum-source terms.
These include an inertia field that offsets momentum variation, a pressure-gradient field, and an Ohmic field that balances collisional drag~\cite{lafleur2014electron,zheng2019enhancement,vass2020electron}.
Dotting each effective field with the conduction current density \(\boldsymbol J_{\mathrm e}\), substituting the expressions for momentum density and flux into Eq.~\eqref{eq:TransportEq}, multiplying by the drift velocity, and rearranging yield the standard BTA form,
\begin{widetext}
\begin{equation}\label{eq:BoltzmannTermAnalysis}
\newcommand{\UBref}{\left(\dfrac{\partial(\rho_{\mathrm m}u)}{\partial t}\right)}
\newcommand{\ub}[2]{\underbrace{\smash{\displaystyle #1}\vphantom{\UBref}}_{#2}}
\boldsymbol J \cdot \boldsymbol E
= \ub{\boldsymbol u \cdot \frac{\partial (\rho_{\mathrm m}\boldsymbol u)}{\partial t}
+ \boldsymbol u \cdot \nabla\!\cdot\!\big(\rho_{\mathrm m}\,\boldsymbol u \boldsymbol u\big)}{P_{\mathrm{in}}}
\;\ub{\mathrel{-}\,\boldsymbol u \cdot (\nabla\!\cdot\!\boldsymbol{\mathcal{P}})}{P_{\mathrm{press}}}
\;\ub{\mathrel{-}\,\boldsymbol u \cdot \left( \frac{\delta \boldsymbol{\rho}_\mathrm{p}}{\delta t} \right)_{\mathrm{coll}} }{P_{\mathrm{Ohmic}}}\,.
\end{equation}
\end{widetext}
The three labeled terms \(P_{\mathrm{in}}\), \(P_{\mathrm{press}}\), and \(P_{\mathrm{Ohmic}}\) are used to identify power-absorption densities associated with distinct mechanisms.
A detailed BTA study for argon CCPs can be found in Ref.~\cite{schulze2018spatio}.

Kinetic-energy transport analysis is an equivalent rearrangement of BTA.
Both come from the same moment hierarchy of the Boltzmann equation, but they emphasize different viewpoints.
The inertial term in BTA can be rewritten as \(P_{\mathrm{in}} = \partial \rho_{\varepsilon, \mathrm k} / \partial t +\nabla\cdot(\rho_{\varepsilon,\mathrm k}\boldsymbol u) +\frac12 \left( \delta \rho_\mathrm{m} / \delta t \right)_{\mathrm{coll}} u^2\).
The first two contributions track the temporal and spatial evolution of the kinetic-energy density \(\rho_{\varepsilon,\mathrm k}\).
The last contribution represents the energy required to accelerate newly created electrons, for example from ionization, up to the drift velocity.
Different rearrangements can introduce sign differences, but the underlying expressions and numerical values are the same.

In typical CCP applications, the dominant contributions in BTA arise from the pressure term \(\boldsymbol u\cdot(\nabla\!\cdot\!\boldsymbol{\mathcal P})\) and the Ohmic term \(\boldsymbol u\cdot(\delta\boldsymbol\rho_{\mathrm p}/\delta t)_{\mathrm{coll}}\).
They are commonly interpreted as collisionless (stochastic) and collisional (Ohmic) power-absorption contributions~\cite{lafleur2014electron}.
Both can be written as an effective force dotted with velocity and have units of power density.
They quantify power absorption associated with pressure forces produced by density and temperature gradients and with effective frictional drag.
The pressure power absorption \(P_{\mathrm{press}}=-\boldsymbol u\cdot(\nabla\!\cdot\!\boldsymbol{\mathcal P})\) describes work done by a net pressure force in a small volume.
It can be positive, converting field energy into kinetic energy, or negative, returning kinetic energy to the field.
Mathematically, this term equals the difference between the pressure-work flux divergence and the pressure--strain term in kinetic-energy transport analysis:
\(\boldsymbol u\cdot(\nabla\!\cdot\!\boldsymbol{\mathcal P})\equiv\nabla\cdot(\boldsymbol{\mathcal P}\cdot\boldsymbol u)-(\boldsymbol{\mathcal P}\cdot\nabla)\cdot\boldsymbol u\).
The volume-integrated form is
\(\int_V \boldsymbol u\cdot(\nabla\!\cdot\!\boldsymbol{\mathcal P})\,\mathrm dV=\oint_{\partial V}(\boldsymbol{\mathcal P}\cdot\boldsymbol u)\cdot\boldsymbol n\,\mathrm dS-\int_V (\boldsymbol{\mathcal P}\cdot\nabla)\cdot\boldsymbol u\,\mathrm dV\).
This shows that pressure-driven power absorption is, to a large extent, converted locally into thermal energy through pressure--strain interaction, or conversely converted back into kinetic energy, while the remaining kinetic energy is transported by pressure work.
The Ohmic term has the same mathematical content in both analyses but is written with opposite sign conventions.
In BTA it is defined as the collisional power absorption \(P_{\mathrm{Ohmic}}=-\boldsymbol u\cdot(\delta\boldsymbol\rho_{\mathrm p}/\delta t)_{\mathrm{coll}}\), whereas in kinetic-energy transport analysis \(\boldsymbol u\cdot(\delta\boldsymbol\rho_{\mathrm p}/\delta t)_{\mathrm{coll}}\) appears as a collisional thermalization term, that is, a loss term (sink) for electron kinetic energy.
This apparent difference reflects two roles of collisions.
Without collisions, electron motion remains phase coherent with a periodic field and the time-averaged power absorption is zero.
Collisions break this coherence and allow net power absorption over an RF period~\cite{lieberman2005principles,pascal11_physic}.
Simultaneously, collisions randomize momentum, reduce anisotropy, and convert directed kinetic energy into thermal energy.
Under these conditions, \(\boldsymbol u\cdot(\delta\boldsymbol\rho_{\mathrm p}/\delta t)_{\mathrm{coll}}\) is typically negative, which indicates kinetic-energy loss, and its negative gives the positive Ohmic absorption density.

\subsection{Pressure--Strain Decomposition and Interpretation}
\label{subsec:pressureStrain}

The present results show that pressure--strain interaction \((\boldsymbol{\mathcal P}\cdot\nabla)\cdot\boldsymbol u\) is also important for kinetic-to-thermal conversion in low-pressure CCPs.
For collisionless plasma systems with periodic boundaries, net change of energy density is determined by \(\boldsymbol J\cdot\boldsymbol E\) and pressure--strain interaction because collisional terms vanish and divergence terms integrate to zero.
In such systems, the pressure--strain interaction provides the only pathway for converting directed kinetic energy into thermal energy~\cite{cassak2022pressure,cassak2022pressureII,yang2017energy_PRE,yang2017energy,du2020energy,barbhuiya2024higher}.

\begin{figure}[htbp]
\centering
\includegraphics[width=0.8\textwidth]{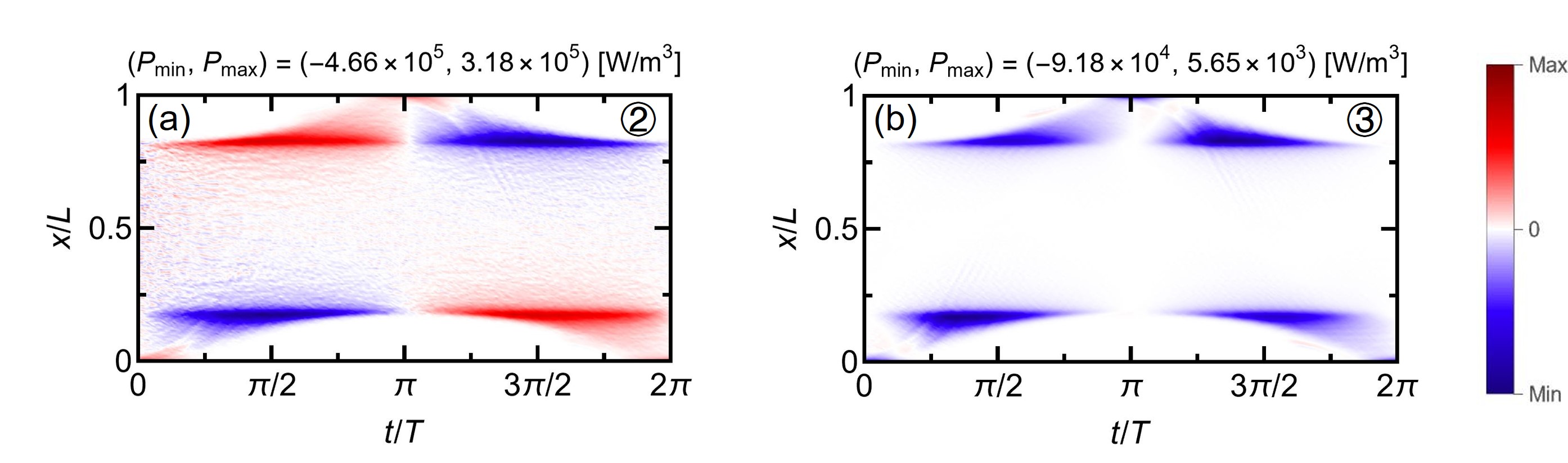}
\caption{\label{fig:strain}Spatiotemporal distributions of individual split terms of pressure--strain interaction corresponding to equation~\eqref{eq:StressStrainInteraction} for electrons.}
\end{figure}

\begin{figure}[htbp]
\centering
\includegraphics[width=0.8\textwidth]{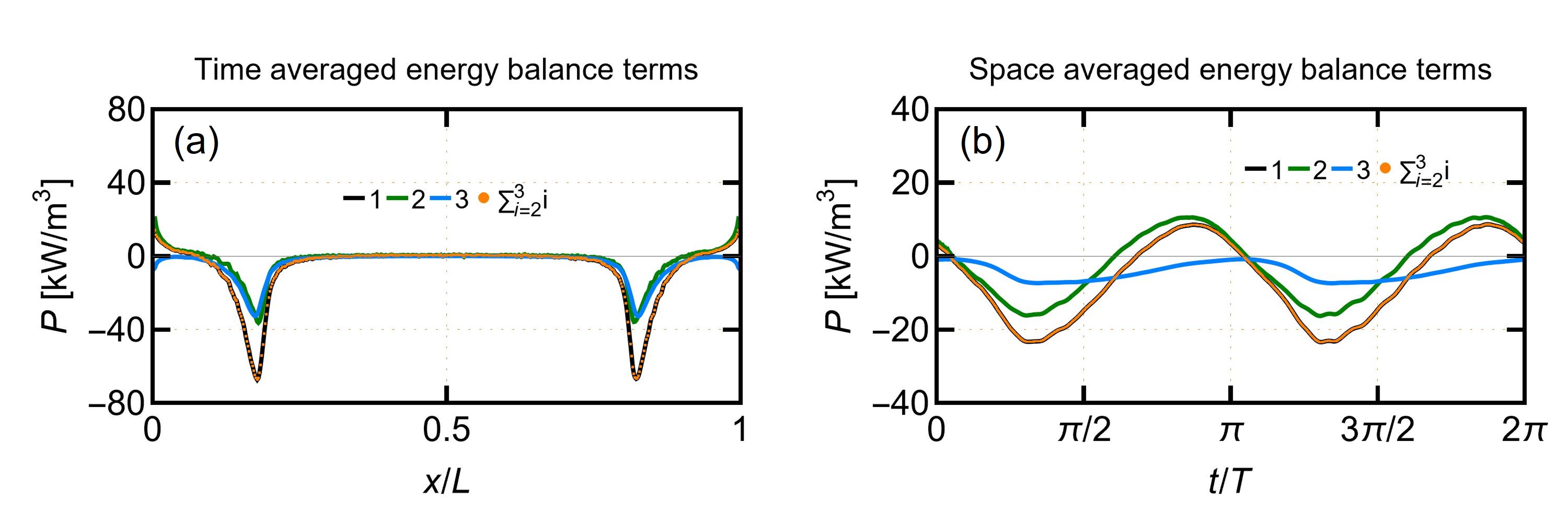}
\caption{\label{fig:strainAvg}Time-averaged spatial profile (a) and space-averaged temporal evolution (b) of pressure--strain interaction and its individual split terms for electrons.
	         The numbered labels in the figure correspond to the terms in equation~\eqref{eq:StressStrainInteraction}.}
\end{figure}

In kinetic and fluid models that capture anisotropy and nonequilibrium, the pressure tensor can be decomposed as
\begin{equation}
\label{eq:PressureTensorDecomposition}
\boldsymbol{\mathcal P}_{ij}=p\,\delta_{ij}+\Pi_{ij}\,,
\end{equation}
where \(p\) is the isotropic pressure, \(\delta_{ij}\) is the Kronecker delta, and \(\Pi_{ij}\) is the deviatoric pressure tensor that measures the departure from isotropy~\cite{yang2017energy_PRE,yang2017energy,yamashita2023inertial}.
In the hydrodynamic limit (fully collisional flow), the Knudsen number \(\mathrm{Kn}=\lambda_{\mathrm{mfp}}/L\) approaches zero, collisions relax the distribution toward a local Maxwellian, and the pressure tensor is nearly isotropic at leading order.
Here, \(\lambda_{\mathrm{mfp}}\) is the electron mean free path and \(L\) is a characteristic macroscopic length scale.
In this limit, \(\Pi_{ij}\) is an \(O(\mathrm{Kn})\) nonequilibrium correction and can be neglected, which corresponds to isotropic fluid descriptions such as the Euler equations.
As collisions become less frequent and \(\mathrm{Kn}\) increases, \(\Pi_{ij}\) can no longer be ignored.
Through the Chapman--Enskog expansion, \(\Pi_{ij}\) can be expressed as a viscous stress proportional to velocity gradients~\cite{vincenti1982introduction}, leading to the Navier--Stokes equations with viscous dissipation.
At even larger \(\mathrm{Kn}\), closures based on local gradients, such as Navier--Stokes/Fourier closures, lose validity and closure must return to kinetic theory.
As shown below, in low-pressure CCPs \(\Pi_{ij}\) can be substantial and reflects nonequilibrium kinetic effects.
With the decomposition in Eq.~\eqref{eq:PressureTensorDecomposition}, the pressure--strain interaction can be written rigorously as
\begin{equation}
\begin{aligned}\label{eq:StressStrainInteraction}
\underbrace{(\boldsymbol{\mathcal P}\!\cdot\!\nabla)\!\cdot\!\boldsymbol u}_{1}
&=p\,\delta_{ij}\,\partial_j u_i+(\mathcal P_{ij}-p\,\delta_{ij})\,\partial_j u_i\\
&=\underbrace{p\,(\nabla\!\cdot\!\boldsymbol u)}_{2}
\;+\;
\underbrace{\boldsymbol{\Pi}:\boldsymbol D}_{3}\,,
\end{aligned}
\end{equation}
where \(D_{ij}=\frac12(\partial_i u_j+\partial_j u_i)-\frac13\delta_{ij}(\nabla\!\cdot\!\boldsymbol u)\) is the traceless strain-rate tensor and \(\nabla\!\cdot\!\boldsymbol u\) is the relative volume change rate of a fluid element~\cite{yang2017energy_PRE,yang2017energy}.
The first term \(p(\nabla\!\cdot\!\boldsymbol u)\) is often called pressure dilatation.
It is widely used in studies of compressible magnetohydrodynamic (MHD) turbulence and describes reversible conversion induced by compressibility~\cite{adhikari2025helmholtz,yang2021energy,hellinger2021scale}.
The second term \(\boldsymbol\Pi:\boldsymbol D=\Pi_{ij}D_{ij}\) arises from the interaction between anisotropic pressure and deformation (double contraction).
It generalizes viscous dissipation in collisional fluids to collisionless kinetic systems and is a key feature that distinguishes kinetic plasmas from classical MHD~\cite{del2016pressure,cassak2022pressure,yang2022pressure,du2020energy,adhikari2025revisiting}.
Figures~\ref{fig:strain} and~\ref{fig:strainAvg} show spatiotemporal distributions and the corresponding time- and space-averaged profiles.

\onecolumngrid

\twocolumngrid

After decomposing pressure--strain interaction into pressure dilatation and viscous-like dissipation, Figs.~\ref{fig:strain} and~\ref{fig:strainAvg} show a clear difference in physical behavior.
Pressure dilatation takes both positive and negative values.
During sheath expansion, the electron fluid is pushed toward the bulk and is locally compressed, \(\nabla\!\cdot\!\boldsymbol u<0\).
Kinetic energy is converted into thermal energy and \(p(\nabla\!\cdot\!\boldsymbol u)<0\), which corresponds to compressional heating.
During sheath collapse, the process reverses.
Fluid expansion converts thermal energy back into kinetic energy, which corresponds to expansion cooling.
Pressure dilatation is therefore reversible.
Its net contribution depends on the symmetry of sheath motion and is small under the present conditions.
In contrast, viscous-like dissipation \(\boldsymbol\Pi:\boldsymbol D\), driven by velocity-space anisotropy, remains negative throughout the sheath.
It acts as an internal friction and provides an irreversible channel that converts directed kinetic energy into disordered thermal energy.
It is the dominant thermalization channel regardless of expansion or collapse.

The kinetic-energy transport analysis reveals a multistage conversion pathway from electromagnetic energy to directed kinetic energy and then to random thermal energy.
A key feature is that the region where electrons gain directed kinetic energy overlaps strongly with the region where kinetic energy is converted into thermal energy.
This indicates that directed kinetic energy is thermalized rapidly and locally near the sheath through pressure--strain interaction and collisions, rather than being transported into the bulk as kinetic energy.
Local thermalization, however, does not imply global equilibration of the electron energy distribution function (EEDF).
Electron temperature can differ substantially between the sheath and the bulk.
These gradients drive subsequent thermal-energy transport and shape the global energy balance.

As commonly recognized, collisions play an essential role in this conversion process.
Momentum scattering in electron--neutral collisions transfers directed kinetic energy into random thermal energy.
At the same time, collisions break field--electron phase coherence and allow net energy gain from a periodic RF field over an RF period~\cite{lieberman2005principles,pascal11_physic}.
Meanwhile, pressure--strain interaction provides another conversion pathway.
Compression, expansion, and deformation can convert kinetic energy into thermal energy even without collisions and can become the dominant thermalization mechanism at low pressure.
Similar behavior has been confirmed in collisionless plasma systems using kinetic methods~\cite{cassak2022pressure,yang2017energy_PRE,yang2017energy,du2020energy,barbhuiya2024higher}.

Finally, electron fluid kinetic energy is only a very small fraction of total electron energy.
At low pressure, directed kinetic energy is generated and transported mainly within the sheath and is almost entirely converted into thermal energy there.
Energy redistribution in the discharge therefore relies mainly on transport of thermal energy.
This thermal-energy transport process is often overlooked in previous studies.

\subsection{Thermal Energy Transport Analysis}
\label{sec:thermalTransport}

We now analyze transport of fluid thermal energy, that is, the internal energy.
We again move source terms to the left-hand side and write the thermal-energy transport equation as
\begin{widetext}
\begin{equation}\label{eq:ThermalTransportEq}
\underbrace{%
\mathrel{-}\,(\boldsymbol{\mathcal P}\!\cdot\!\nabla)\!\cdot\!\boldsymbol u
\;+\; \frac12\!\left(\frac{\delta \rho_{\mathrm m}}{\delta t}\right)_{\mathrm{coll}}\,u^{2}
\;-\; \boldsymbol u\!\cdot\!\left(\frac{\delta \boldsymbol{\rho}_{\mathrm p}}{\delta t}\right)_{\mathrm{coll}}%
}_{1}
=
\underbrace{\frac{\partial \rho_{\varepsilon,\mathrm{th}}}{\partial t}}_{2}
\;+\;
\underbrace{\nabla\!\cdot\!\big(\rho_{\varepsilon,\mathrm{th}}\,\boldsymbol u\big)}_{3}
\;+\;
\underbrace{\nabla\!\cdot\!\boldsymbol q}_{4}
\;\underbrace{\mathrel{-}\,\left(\frac{\delta \rho_\varepsilon}{\delta t}\right)_{\mathrm{coll}}}_{5}\,.
\end{equation}
\end{widetext}
Here, \(\rho_{\varepsilon,\mathrm{th}}=\frac32 p\) is the thermal energy density and \(\boldsymbol q\) is the heat flux.
The terms have the following meanings:
\begin{enumerate}[leftmargin=*]
\item Term 1 is the thermal-energy source due to conversion from fluid kinetic energy.
      It is the negative of terms 2--4 on the left-hand side of Eq.~\eqref{eq:KineticTransportEq}.
      It describes how directed kinetic energy gained after field acceleration is converted into thermal energy through pressure--strain interaction and collisional processes.
\item \(\partial\rho_{\varepsilon,\mathrm{th}}/\partial t\) is the local time variation of thermal energy density.
\item \(\nabla\!\cdot(\rho_{\varepsilon,\mathrm{th}}\boldsymbol u)\) is convective transport of thermal energy by bulk drift.
\item \(\nabla\!\cdot\boldsymbol q\) is the divergence of microscopic heat flux and represents non-convective transport driven by thermal motion.
\item \(-(\delta\rho_\varepsilon/\delta t)_{\mathrm{coll}}\) is collisional loss of thermal energy.
      In the present system without super-elastic collisions, it is positive and represents true loss of thermal energy and total energy.
\end{enumerate}

\begin{figure*}[htbp]
\centering
\includegraphics[width=0.8\textwidth]{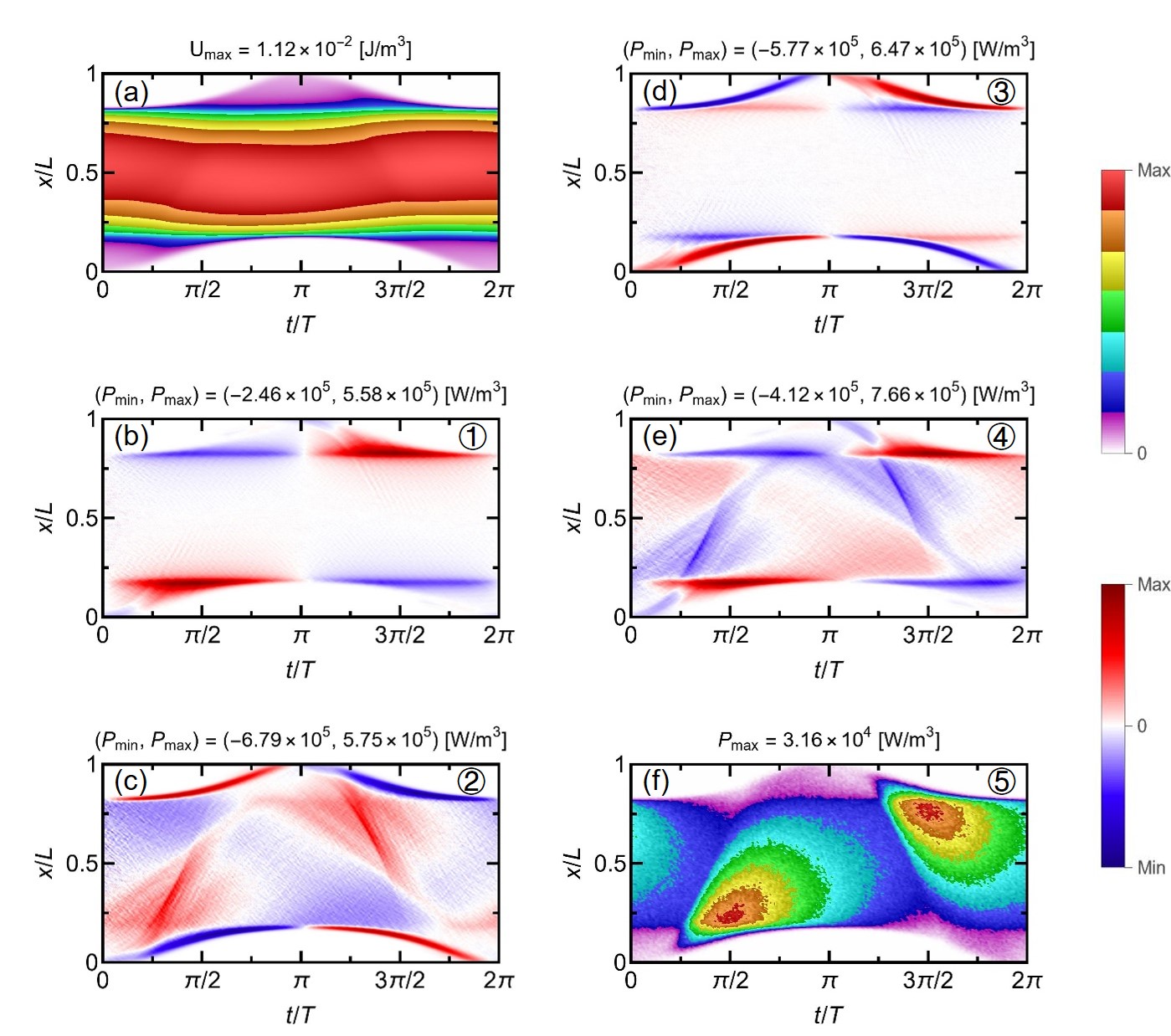}
\caption{\label{fig:thermal}Spatiotemporal distributions of thermal energy density \(\rho_{\varepsilon, \mathrm{th}}\) (a) and each transport term (b-f) corresponding to the thermal energy transport equation~\eqref{eq:ThermalTransportEq} for electrons.}
\end{figure*}

\begin{figure*}[htbp]
\centering
\includegraphics[width=0.8\textwidth]{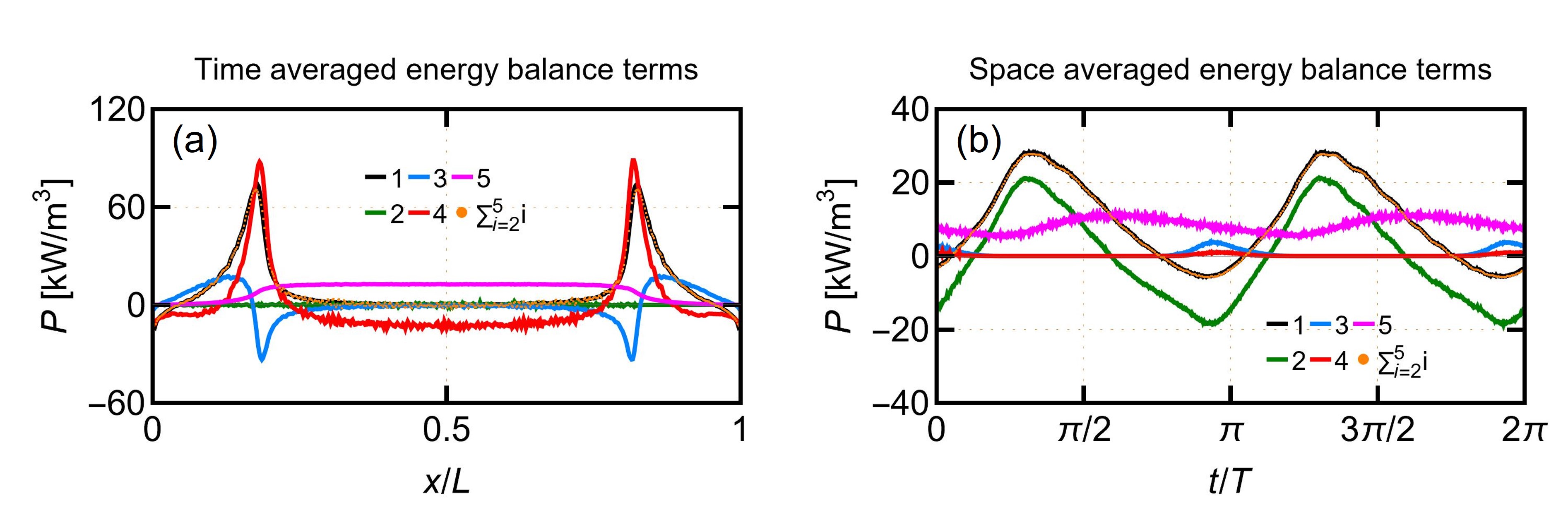}
\caption{\label{fig:thermalAvg}Time-averaged spatial profile (a) and space-averaged temporal evolution (b) of thermal energy transport terms for electrons.
         The numbered labels in the figure correspond to the terms in the thermal energy transport equation~\eqref{eq:ThermalTransportEq}.}
\end{figure*}

Figure~\ref{fig:thermal} shows spatiotemporal distributions of thermal energy density and each term in Eq.~\eqref{eq:ThermalTransportEq}.
Figure~\ref{fig:thermalAvg} shows the time-averaged spatial profile and space-averaged temporal evolution.
The close agreement of the sums on both sides of Eq.~\eqref{eq:ThermalTransportEq} verifies energy conservation in the diagnostics.

As shown in Fig.~\ref{fig:thermal}(a), electron thermal energy accounts for the vast majority of total electron energy in CCPs.
Its spatiotemporal distribution is therefore similar to that of total energy and is mainly concentrated in the bulk.
From the definition of thermal energy, an increase in thermal energy density can be caused by an increase in electron temperature, which is actual heating, and by an increase in electron density, which corresponds to compression.
Both can contribute.
As discussed in Subsection~\ref{sec:kineticTransport}, conversion of kinetic energy provides the thermal-energy source term in Fig.~\ref{fig:thermal}(b).
This source is dominated by pressure--strain interaction and collisions in the sheath and is nearly zero in the bulk.
Figure~\ref{fig:thermalAvg} shows that the time-variation term reaches a periodic balance at each position, while its space-averaged temporal evolution oscillates with a phase similar to the source term.
This indicates that thermal energy is modulated by the RF field rather than growing or decaying monotonically.

Figure~\ref{fig:thermal}(d) shows that the convective thermal-energy transport term is positive near the electrode side and negative toward the bulk side.
This indicates net transport of thermal energy from the sheath toward the bulk.
Its distribution resembles that of the pressure-work transport term \(\nabla\!\cdot(\boldsymbol{\mathcal P}\cdot\boldsymbol u)\) in the kinetic-energy equation and is confined to the sheath.
Both kinetic and thermal energy are transported by the moving charged particles, so consistency between these terms indicates that electrons carry both forms during convective transport in the sheath.

The key term that transfers thermal energy from the sheath into the bulk is the heat-flux divergence \(\nabla\!\cdot\boldsymbol q\).
As shown in Fig.~\ref{fig:thermal}(e), it spans the entire discharge domain and is the only channel that transports electron energy into the bulk.
The temporal evolution in Fig.~\ref{fig:thermalAvg}(b) shows that heat-flux divergence produces only a small boundary loss during sheath collapse, and its magnitude is much smaller than that of convection.
The main loss of electron thermal energy, and therefore of total electron energy, is inelastic collisions, represented by the collisional loss term.
Its time-averaged profile shows a stable positive value in the bulk that balances the heat-flux divergence.
This indicates that thermal energy transported from the sheath by microscopic heat flux is dissipated in the bulk by inelastic collisions.
This is consistent with total energy transport because directed kinetic energy is mainly confined near the sheath edge and the bulk energy is predominantly thermal.

To better understand spatial transport of thermal and total energy, we examine the energy flux.
As discussed above, the energy flux can be decomposed as
\begin{equation}\label{eq:EnergyFlux}
\underbrace{\boldsymbol\Gamma_\varepsilon}_{1}
=
\underbrace{\rho_\varepsilon\boldsymbol u}_{2}
+
\underbrace{\boldsymbol{\mathcal P}\cdot\boldsymbol u}_{3}
+
\underbrace{\boldsymbol q}_{4}\,.
\end{equation}
This shows that the energy flux can be written as the sum of three contributions.
The second term is convective transport of total energy density by bulk drift.
The third term is the energy flux associated with pressure work.
The fourth term is non-convective transport due to random thermal motion~\cite{bittencourt2013fundamentals}.
The heat flux is defined as \(\boldsymbol q=\frac12\rho_{\mathrm m}\langle w^2\boldsymbol w\rangle\), where \(\rho_{\mathrm m}\) is the mass density and \(\boldsymbol w=\boldsymbol v-\boldsymbol u\) is the peculiar velocity, i.e., the thermal velocity relative to the mean flow.
If the distribution is even (symmetric) in \(\boldsymbol w\) in the local rest frame, that is, \(f(\boldsymbol w)=f(-\boldsymbol w)\), for example in an isotropic local equilibrium, the integrand is odd (antisymmetric) and \(\boldsymbol q=0\).
The heat flux therefore measures third-moment asymmetry, or skewness, relative to local equilibrium, and it is closely related to nonequilibrium transport such as heat conduction.

\begin{figure}[htbp]
\centering
\includegraphics[width=0.8\textwidth]{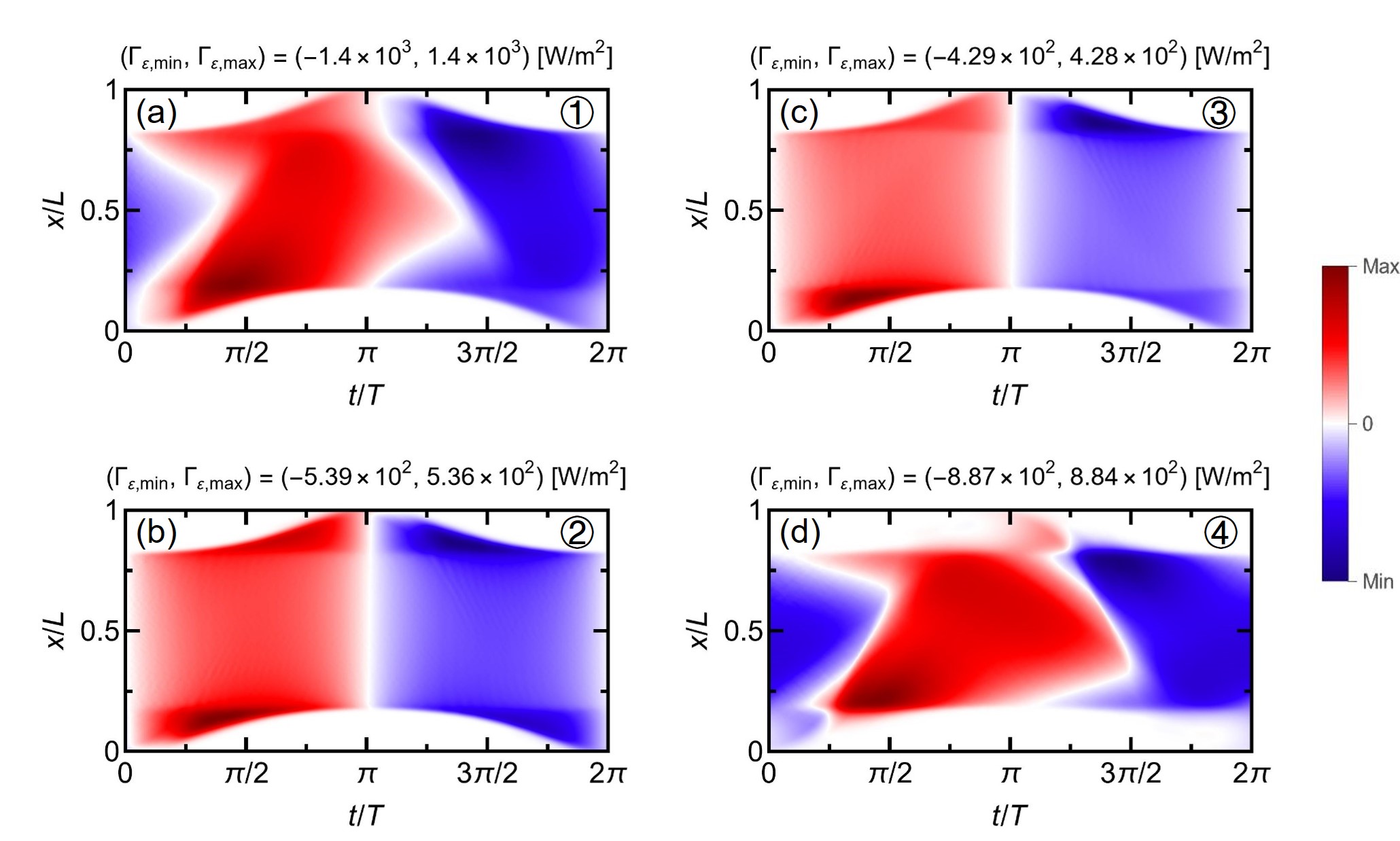}
\caption{\label{fig:energyFlux}Spatiotemporal distributions of energy flux (a) and its individual split terms (b-d) corresponding to equation~\eqref{eq:EnergyFlux} for electrons.}
\end{figure}

\begin{figure}[htbp]
\centering
\includegraphics[width=0.8\textwidth]{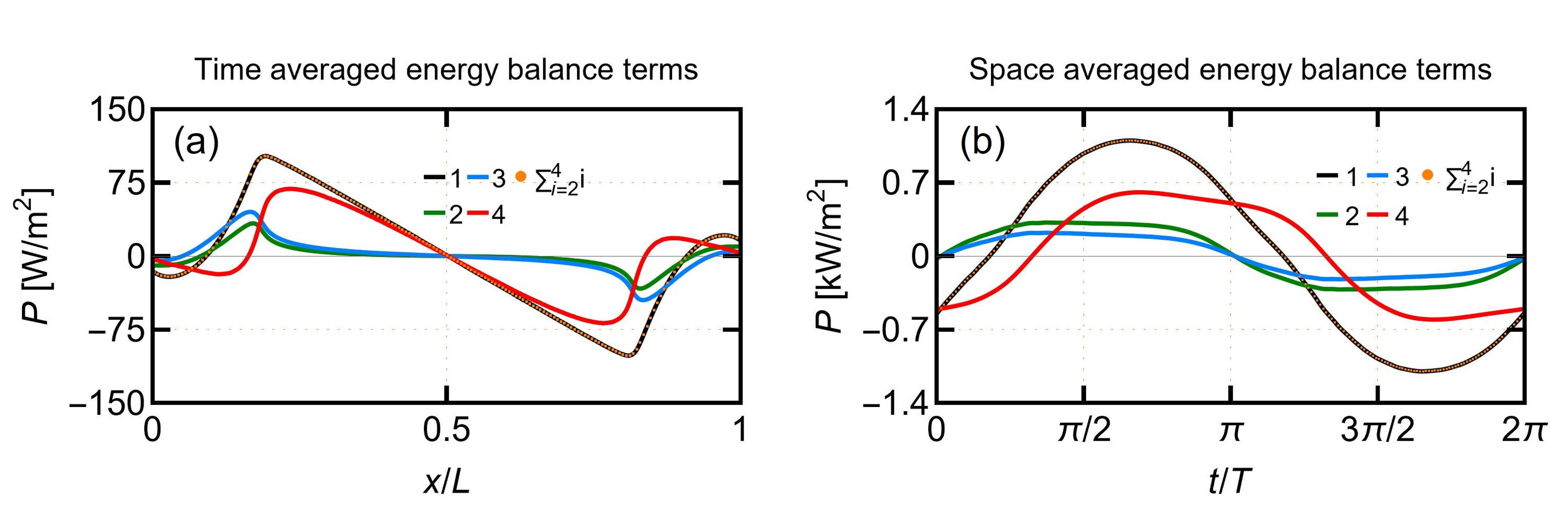}
\caption{\label{fig:energyFluxAvg}Time-averaged spatial profile (a) and space-averaged temporal evolution (b) of energy flux and its individual split terms for electrons.
         The numbered labels in the figure correspond to the terms in equation~\eqref{eq:EnergyFlux}.}
\end{figure}

Figures~\ref{fig:energyFlux} and~\ref{fig:energyFluxAvg} show spatiotemporal distributions of each term in Eq.~\eqref{eq:EnergyFlux} and the corresponding averaged results.

\onecolumngrid

\twocolumngrid

The spatial profiles in Fig.~\ref{fig:energyFluxAvg}(a) show that the convective term and the pressure-work term are concentrated in the sheath because the electron flow velocity \(\boldsymbol u\) is much larger there than in the bulk.
In the bulk, spatial redistribution of energy is dominated by the microscopic heat flux.
This is consistent with the earlier analysis of divergence terms in kinetic- and thermal-energy transport equations.

In classical fluid transport theory, the electron heat flux is often closed by Fourier's law \(\boldsymbol q_{\mathrm e}=-\kappa_{\mathrm e}\nabla T_{\mathrm e}\), where \(\kappa_{\mathrm e}\) is the electron thermal conductivity and \(T_{\mathrm e}\) is the electron temperature.
This relation describes diffusive transport of electron thermal energy from hot to cold regions, and \(\kappa_{\mathrm e}\) depends on parameters such as plasma density and collision frequency.
Qualitatively, this resembles the present low-pressure CCP.
RF power is absorbed in the sheath and creates a high-temperature region, while the bulk is relatively cold.
A temperature gradient would then drive a heat flux.
In our simulation, however, the heat flux vector differs strongly from the electron temperature gradient.
The temperature gradient is mainly confined within the maximum sheath width and is nearly zero in the bulk (data not shown), while the heat flux extends across sheath and bulk.
The reason is that Fourier's law is a local approximation based on the second velocity moment, the electron temperature.
It assumes that heat flux is driven by the temperature gradient, which reflects the variance of the velocity distribution.
However, the electron heat flux intrinsically depends on the third velocity moment \(\langle w^2\boldsymbol w\rangle\) and reflects skewness of the distribution.
It therefore emphasizes net energy flow caused by velocity-space asymmetry, especially directional propagation of energetic electrons, rather than a purely diffusive process driven by local temperature gradients.

At low pressure, the electron mean free path exceeds the reactor scale and the temperature-gradient scale.
Energy gain and transport are strongly nonlocal.
Electrons can gain energy from sheaths far from their current location and propagate nearly collisionlessly, which produces an EEDF with a high-energy tail.
High-energy electrons and low-energy electrons then have different transport behaviors.
Low-energy electrons dominate in number density and mainly oscillate locally with the instantaneous electric field, so the electron flux varies approximately sinusoidally in time~\cite{wilczek2020electron}.
This is reflected in the convective and pressure-work components, which alternate in sign over the RF period in Fig.~\ref{fig:energyFlux}.
High-energy electrons are fewer in number but can carry a disproportionately large share of the energy flux.
Their directional propagation can be masked in the mass flux, but it dominates the microscopic heat flux and is crucial for energy transport.
Under these conditions, \(\boldsymbol q_{\mathrm e}(x)\) does not represent diffusive heat conduction driven by \(\nabla T_{\mathrm e}(x)\).
It is a nonlocal energy transport dominated by streaming energetic electrons.
This also explains the time evolution in Fig.~\ref{fig:energyFluxAvg}(b).
The microscopic heat flux shows a phase lag relative to other components and largely determines the phase of the total energy flux.

In summary, the kinetic nature of the electron heat flux clarifies how energy is transported from the sheath, where it is deposited, to the bulk, where it is dissipated, in low-collisionality CCPs.
Because the heat flux is a third-order moment, Fourier-law closures can fail in strongly nonlocal regimes, which motivates higher-order nonlocal closures or hybrid kinetic--fluid strategies.

\section{CONCLUSIONS}
\label{sec:conclusions}

We have developed a kinetic--moment analysis framework based on PIC/MCC simulations and the first three velocity moments of the Boltzmann equation.
The framework provides a self-consistent diagnosis of electron energy dynamics in low-pressure CCPs and clarifies how absorption, conversion, transport, and dissipation are coupled.
By decoupling total energy into directed kinetic energy and random thermal energy, we identify the conversion pathway among electromagnetic energy, electron kinetic energy, and electron thermal energy and quantify spatial redistribution.

In the present low-pressure regime, electron energy absorption occurs mainly in the sheath, where electrons are accelerated and gain directed kinetic energy, which produces an anisotropic velocity distribution.
Directed kinetic energy is then converted locally into thermal energy through pressure--strain interaction and collisions that are dominated by elastic scattering.
The thermal energy is transported toward the bulk by microscopic heat flux.
It is ultimately dissipated in the bulk through inelastic collisions such as excitation and ionization.
This dissipation sustains global particle and energy balance.

At low pressure, the mean free path is comparable to the reactor size, so energy absorption in the sheath and dissipation in the bulk are separated in space and time, while kinetic-to-thermal conversion remains localized near the sheath edge.
This shows that rapid local thermalization can coexist with strongly nonlocal transport.
Under the present conditions, pressure--strain interaction induced by spatially nonuniform flow is the dominant conversion mechanism, in addition to irreversible collisions.
The pressure--strain term can be separated into reversible pressure dilatation and irreversible viscous-like dissipation.
The net periodic effect prevents a full return of the energy supplied by the RF electric field and leads to net electron power absorption.
The pressure--strain interaction is distinct from collision-driven Ohmic heating and may provide a deeper, first-principles physical picture of the so-called collisionless or stochastic heating.

Our results also show that, in low-pressure CCPs with strong nonlocal effects, anisotropic components of the pressure tensor play a key role in kinetic-to-thermal conversion and pressure cannot be treated as a scalar.
Spatial transport of electron thermal energy and total energy depends critically on microscopic heat flux.
This heat flux deviates strongly from Fourier's law and its nonlocality is essential for spatial energy redistribution.
This helps explain why traditional fluid simulations fail at low pressure.

Because the present framework diagnoses macroscopic quantities directly from PIC/MCC data and reconstructs moment equations without closure assumptions, it provides a solid basis for quantitative studies of energy balance and thermalization.
It can also serve as a reliable benchmark for fluid and hybrid models.
Finally, while the present transport analysis relies on ensemble averaging and can smooth out some microscopic details of energetic electrons, the kinetic--moment framework has substantial potential for extension.
Future work may combine the present approach with Lagrangian particle tracking to complement the fluid viewpoint.
It can also be extended to multi-dimensional geometries, magnetized plasmas, and more complex chemistry, and it can support high-fidelity plasma process modeling.

\section*{ACKNOWLEDGMENTS}
This work is partly supported by the National Natural Science Foundation of China (Grant No. 11975047).

\section*{AUTHOR DECLARATIONS}
\subsection*{Conflict of Interest}
The authors have no conflicts to disclose.

\bibliography{references.bib}
\end{document}